\documentclass[12pt,english]{article}
\usepackage{lmodern}
\usepackage[T1]{fontenc}
\usepackage[latin9]{inputenc}
\usepackage{amsmath}
\usepackage{amssymb}
\usepackage{esint}

\usepackage{geometry}
\makeatletter
\usepackage{float} 
\usepackage{hyperref}
\usepackage{cite}
\usepackage{babel}
\usepackage{mathdots}
\usepackage{amsfonts}
\usepackage{mathrsfs}
\usepackage{amsmath}
\usepackage{esint}
\usepackage{graphicx}
\usepackage{youngtab}
\usepackage{multicol}
\usepackage{multirow}
\usepackage{slashed}
\usepackage{bm}
\usepackage{relsize}
\usepackage[dvipsnames,table,xcdraw]{xcolor}
\usepackage{subfigure}
\usepackage{feynmp}
\usepackage{cancel}
\allowdisplaybreaks

\makeatother

\usepackage[dvipsnames]{xcolor}

\begin{document}
\author{Wan-Zhe Feng\footnote{Email: vicf@tju.edu.cn} ~and Zi-Hui Zhang\footnote{Email: zhangzh\_@tju.edu.cn}\\
\textit{\small{Center for Joint Quantum Studies and Department of Physics,}}\\
\textit{\small{School of Science, Tianjin University, Tianjin 300350, PR. China}}}
\title{Darker matter generating from the dark}
\date{}

\maketitle

\begin{abstract}
  The non-detection of dark matter may be attributed to the dark matter residing in a darker hidden sector.
  We explore the possibility that a hidden sector
  produced through the freeze-in mechanism,
  can further generate an even more hidden sector via an additional freeze-in process.
  Such a two-step freeze-in process produces dark matter coupled weaker-than-ultraweakly to the standard model particles,
  and is thus referred to as the ``darker matter''.
  To illustrate the two-step freeze-in process,
  we study a model featuring two $U(1)$ hidden sectors.
  The first $U(1)$ sector is directly coupled to the standard model with feeble interactions,
  while the second $U(1)$ sector is directly coupled to the first $U(1)$ sector
  and thus only indirectly to the standard model, rendering it darker.
  Remarkably, darker matter candidates residing in the second darker $U(1)$ sector,
  generated from the two-step freeze-in process,
  can account for almost the entire observed dark matter relic density.
  The darker matter, interacted with standard model particles through ultraweak couplings,
  can exhibit velocity-dependent self-interacting cross-sections,
  which potentially provides an explanation for addressing problems associated with cosmic small-scale structures.
  Additionally, the dark photon darker matter residing in the darker hidden sector
  can be responsible for the galactic 511 keV photon signal, consistent with various dark matter density profiles.
\end{abstract}
\newpage{}

\tableofcontents{}

\section{Introduction}\label{sec:Intro}

After years of searches for dark matter yielding no findings,
research into dark matter from both theoretical and experimental perspectives has faced significant challenges in recent years.
It is widely believed that dark matter resides in one or multiple hidden sectors beyond the Standard Model (SM),
and thus its non-detection may be due to dark matter existing in an even more hidden sector in the Universe.
In general, the SM can be extended by hidden sectors either directly or indirectly connected to the SM~\cite{Aboubrahim:2021ycj,Aboubrahim:2021dei}.
However,
the literature has not sufficiently explored the possibility of dark matter originating from multiple hidden sectors.
The dynamics of the ``dark'' physics within these hidden sectors may have been underestimated and largely overlooked in the past.
Hence, it is essential to explore interactions across multiple hidden sectors that contain various types of dark matter candidates.
In particular, as was discussed in~\cite{Feng:2024blk},
dark matter may interact more strongly within multiple hidden sectors than with the SM.
Thus, the non-detection of WIMP\footnote{
Traditionally, the term ``WIMP'' is used to denote a massive particle,
typically with a mass around the electroweak scale $\sim \mathcal{O}(100)$~GeV,
coupled to SM particles with a strength comparable to the weak interaction coupling constant.
This type of WIMP serves as a well-motivated dark matter candidate,
as they can be produced thermally in the early universe with an annihilation cross-section of just the required order.
While ``WIMP'' has later been applied to a broader range of particles,
whose masses can span from MeV to TeV range.
Additionally, their couplings with SM particles can be several orders of magnitude lower than the electroweak coupling constant.
As dark matter candidates,
they can still be produced thermally, and their abundance is diminished through the freeze-out mechanism down to the observed value of dark matter relic density.} may be attributed to dark matter
predominantly annihilating into a darker concealed sector rather than into the SM sector.
The WIMP dark matter can either annihilate efficiently with the assistance of a darker sector,
or it will transform into another type of dark matter candidate~\cite{Feng:2024blk}.

Another possible explanation for the non-detection of dark matter is
the dark matter being produced via the freeze-in mechanism~\cite{Hall:2009bx,Cheung:2010gj,Chu:2011be,Bernal:2017kxu,Berger:2018xyd}
\cite{Aboubrahim:2019kpb,Aboubrahim:2020wah,Aboubrahim:2020lnr,Aboubrahim:2021ycj,Aboubrahim:2022bzk,Feng:2023ubl,Feng:2024pab},
where dark particles can only couple to the visible sector via feeble couplings.
In contrast to the freeze-out thermal WIMPs,
freeze-in dark matter is generated from feeble couplings with the visible sector starting from a negligible amount,
rather than depleting into it.
The number densities of dark particles increase gradually as the temperature of the Universe drops down.
Although the freeze-in mechanism is clear and simple,
while the abundance of the dark particle is difficult to calculate
when there are multiple freeze-in particles and they can interact with each other.
A computation method is developed in~\cite{Aboubrahim:2020lnr}
to compute a freeze-in hidden sector with hidden sector interactions.

Since the nature of dark matter remains a mystery,
the pursuit of new and diverse possibilities for its generation and evolution remains an open avenue of research.
Dark particles from multiple hidden sectors may engage in more complex interactions,
leading to intricate evolutions.
Darker hidden sectors, which only indirectly couple to the SM, can provide viable ``darker matter'' candidates,
requiring a deeper investigation into the dark aspects of the Universe.
The computation method developed in~\cite{Aboubrahim:2020lnr}
can be extended to analyze dark interactions among multiple hidden sectors that are produced via the freeze-in mechanism,
though when more freeze-in hidden sectors are present,
the computation difficulty is highly increased~\cite{Aboubrahim:2021ycj,Aboubrahim:2022bzk}.
This approach is crucial for identifying new physics signals in concordance with various experimental constraints
and the dark matter relic density constraint.

In this paper, we explore the possibility that the darker matter can be
almost entirely generated from a frozen-in hidden sector through a further freeze-in process.
Specifically,
we investigate the possibility that a hidden sector, which directly couples to the SM sector through feeble interactions,
can further undergo freeze-in processes to produce a vibrant, self-interacting darker hidden sector.
Such a two-step freeze-in process can generate darker matter candidates
coupled weaker-than-ultraweakly to the SM sector,
which is demonstrated by a two-$U(1)$ model, graphically shown in Fig.~\ref{Fig:2U1Model}.

In the two-$U(1)$ model,
the first $U(1)$ sector mixed with the SM directly through kinetic terms,
while the second $U(1)$ sector mixed with the first $U(1)$ directly
and thus coupled to the SM indirectly.
The first $U(1)$ serves as a portal to further generate
darker particles within the second $U(1)$ sector through the freeze-in mechanism.
The dark matter candidates can be a combination of dark particles from both of the two $U(1)$ sectors.
Remarkably, the darker matter generated from the dark through the two-step freeze-in process
can constitute almost the entire observed dark matter relic density.

The phenomenology of darker matter candidates is rich and intriguing.
The pursuit of self-interacting dark matter particles (SIDM) has been longstanding,
with the potential to address problems associated with cosmic small-scale structures~\cite{Tulin:2013teo,Kaplinghat:2015aga}.
The darker matter, which interacts with SM particles through ultraweak couplings,
can exhibit velocity-dependent self-interacting cross-sections,
and thus provide the explanation for the small-scale structures issues.
The dark photon 
with mass  $2m_e < M_{\gamma^\prime} \lesssim 6~{\rm MeV}$
from the indirectly connected darker hidden sector is a viable dark matter candidate and
its decay with a minuscule amount to the electron and positron pair is capable of explaining the longstanding galactic 511 keV signal~\cite{1st511,511-2,511-3,511-4,511-5,511-6,Knodlseder:2003sv,Jean:2003ci,Weidenspointner:2004my,Knodlseder:2005yq,Siegert:2015knp,Kierans:2019aqz}.

The paper is organized as follows.
In Section~\ref{Sec:MultiHS} we discuss the physics emerging from the dark side of the Universe,
focusing on the interplay among dark particles across multiple hidden sectors,
which may reveal a rich phenomenology.
To illustrate the dynamics of dark physics from multiple hidden sectors,
we present a two-$U(1)$ model beyond the SM, discussing their mixing in Section~\ref{Sec:2U1}.
The evolution of the dark particles from the two hidden sectors are derived in Section~\ref{Sec:Evo2U1}.
The potential of darker matter as SIDM to explain cosmic small-scale structure problems is discussed in
Section~\ref{Sec:SIDM}.
In Section~\ref{Sec:511}, we study the dark photon dark matter interpretation of the galactic 511~keV signal.
The evolution of a multi-temperature universe is reviewed briefly in Appendix~\ref{App:MultiT}.
A full derivation of simultaneous diagonalization of the $4\times 4$ kinetic and mass matrices using the perturbation method is given in Appendix~\ref{App:2U1dia}.
We conclude in Section~\ref{Sec:Con}.

\section{Dark matter candidates from darker hidden sectors}\label{Sec:MultiHS}

In general, a hidden sector may be directly connected to the SM through viable portal interactions. Alternatively, it may remain disconnected from the SM, interacting only indirectly via other hidden sectors serving as mediators.
In this paper, we delve deeper into the physics of multiple hidden sectors and investigate a novel framework in which
a freeze-in generated hidden sector can further freeze-in produce a darker hidden sector.
Dark matter from such darker hidden sector thus couples more weakly to the SM sector,
and is therefore termed  the ``darker matter''.
Unexpectedly, such two-step freeze-in process can surprisingly generate viable dark matter candidates
which constitute almost the entire observed dark matter relic density.
To illustrate such a scenario, we investigate a two-$U(1)$ model where the first $U(1)_1^\prime$ sector
is generated through freeze-in production from SM particles,
while the second $U(1)_2^\prime$ sector arises from its feeble coupling to the first $U(1)_1^\prime$ sector.
The $U(1)_2^\prime$ sector does not couple with the SM sector directly
but through $U(1)_1^\prime$ sector indirectly.
In this section we are going to discuss the details of the two-$U(1)$ model, shown in Fig.~\ref{Fig:2U1Model},
as well as the evolution of all dark particles.

\subsection{The two-$U(1)$ extension of the SM}\label{Sec:2U1}

\begin{figure}
	\begin{center}
		\includegraphics[scale=0.45]{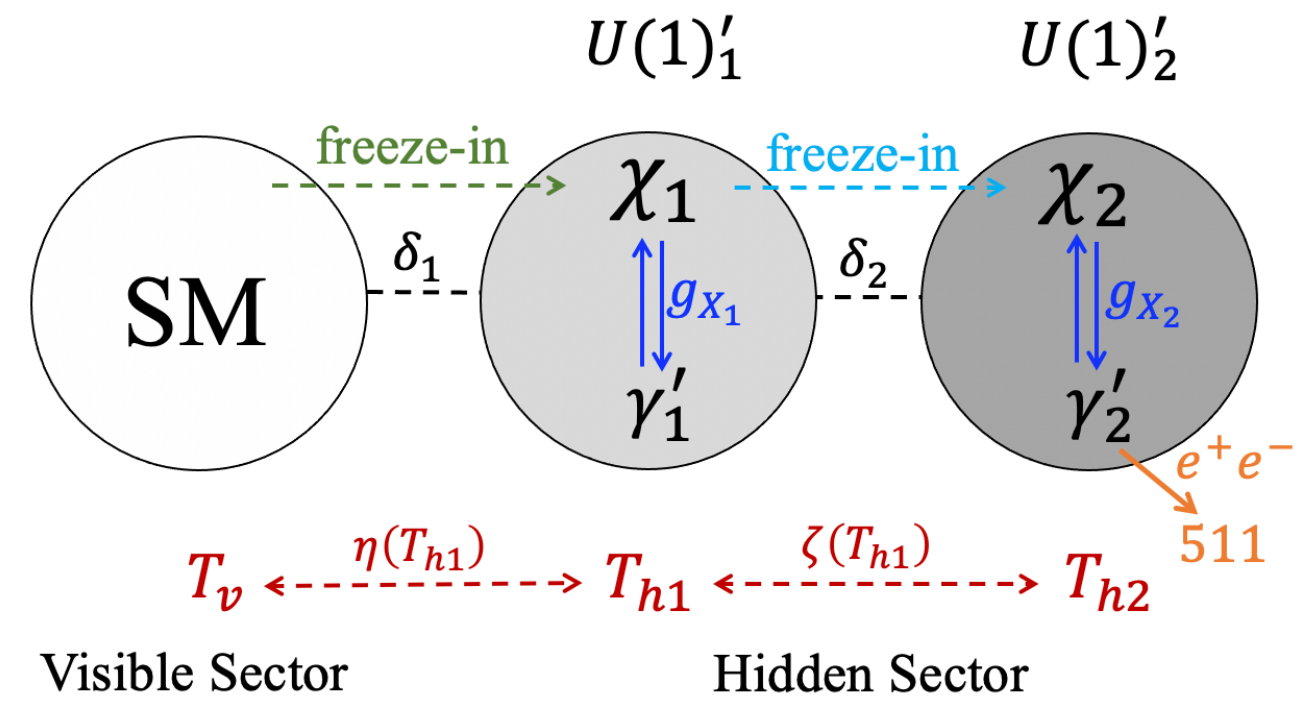}
		\caption{
			A graphic illustration of a general two-$U(1)$ model we discuss.
            The $U(1)_{1}^{\prime}$
			hidden sector connects to the SM and the second hidden sector $U(1)_{2}^{\prime}$
			via kinetic mixing characterized by parameters $\delta_{1}$ and $\delta_{2}$.
            Thus $U(1)_{2}^{\prime}$ couples to the SM indirectly and
            with the strength proportional to $\delta_1\times \delta_2$.
            $U(1)_{1}^{\prime}$ sector particles are produced via freeze-in processes from SM particles
            while $U(1)_{2}^{\prime}$ particles are produced via freeze-in mostly
			from the $U(1)_{1}^{\prime}$ sector.
			Each $U(1)$ hidden sector possesses a different temperature $T_{h1},T_{h2}$ respectively,
            related to the visible sector temperature $T_{v}$ through functions
            $\eta(T_{h1})= T_{v}/T_{h1}$ and $\zeta(T_{h1})= T_{h2}/T_{h1}$.
            The dark matter candidates are a combination of $\chi_{1},\chi_{2},\gamma_{2}^{\prime}$,
            and $\chi_{2},\gamma_{2}^{\prime}$ from the $U(1)_{2}^{\prime}$ sector are the darker matter candidates.
            The dark photon $\gamma^\prime_2$ from the $U(1)_2$ sector
            may potentially account for the galactic 511~keV signal.}
		\label{Fig:2U1Model}
	\end{center}
\end{figure}

The model with two $U(1)$ hidden sectors is graphically shown in Fig.~\ref{Fig:2U1Model}.
In this setup, the first $U(1)_1^{\prime}$ mixed with the hypercharge with kinetic mixing parameter $\delta_1$
and the second $U(1)_2^{\prime}$ mixed with $U(1)_1^{\prime}$ with kinetic mixing parameter $\delta_2$.
Thus the $U(1)_2^{\prime}$ couples to the SM indirectly using $U(1)_1^{\prime}$ as a portal.
The total Lagrangian of the model is given by
\begin{equation}
\mathcal{L}=\mathcal{L}_{{\rm SM}}+\mathcal{L}_{{\rm hid}}+\mathcal{L}_{{\rm mix}}\,,
\end{equation}
where $\mathcal{L}_{{\rm hid}}$ is given by
\begin{align}
\mathcal{L}_{{\rm hid}} & =-\frac{1}{4}F_{1\mu\nu}F_{1}^{\mu\nu}-\frac{1}{4}F_{2\mu\nu}F_{2}^{\mu\nu}
+g_{X_1}\bar{\chi}_1\gamma^{\mu}\chi_1 C_{\mu}
+g_{X_2}\bar{\chi}_2\gamma^{\mu}\chi_2 D_{\mu}  \\
 & \quad-\frac{1}{2}(M_{C}C_{\mu}+\partial_{\mu}\sigma_{1})^{2}-\frac{1}{2}(M_{D}D_{\mu}+\partial_{\mu}\sigma_{2})^{2}
-m_{\chi_1}\bar{\chi}_1\chi_1 -m_{\chi_2}\bar{\chi}_2\chi_2 \nonumber
\end{align}
and $\mathcal{L}_{{\rm mix}}$ includes the kinetic mixing between
$U(1)_{Y}$ and $U(1)_{1}^{\prime}$, and the kinetic mixing between $U(1)_{1}^{\prime}$
and $U(1)_{2}^{\prime}$
\begin{equation}
\mathcal{L}_{{\rm mix}}=-\frac{\delta_{1}}{2}F_{Y\mu\nu}F_{1}^{\mu\nu}-\frac{\delta_{2}}{2}F_{1\mu\nu}F_{2}^{\mu\nu}\,,
\end{equation}
where $\delta_{1},\delta_{2}$ are kinetic mixing parameters.\footnote{Here the kinetic mixing (or any kind of mixings) between the SM and $U(1)_2$ sector is
  intentionally set to be negligible. This choice is motivated by our focus on exploring physics originating from a darker sector that is only indirectly connected to the SM with weaker-than-ultraweak interactions.
  If a kinetic mixing between $U(1)_2$ and the hypercharge does exist, we require the kinetic mixing parameter to satisfy $\lesssim \delta_1\times \delta_2$.}
In our analysis we assume $\delta_{1}\lesssim10^{-10}$ and $\delta_{2}\lesssim10^{-6}$.
The two extra $U(1)$ fields acquire masses through the Stueckelberg mechanism.
In the unitary gauge, one has the $U(1)_{1}^{\prime}$ gauge boson $C_{\mu}$
and $U(1)_{2}^{\prime}$ gauge boson $D_{\mu}$.
After the mixing, the above gauge bosons from the extra $U(1)$ sectors
become $\gamma_1^\prime$ and $\gamma_2^\prime$ in the mass eigenbasis,
which are mostly $C_\mu$ and $D_\mu$ respectively because of the tiny kinetic mixing.
Dark fermions are generically present in the two $U(1)$ hidden sectors:
$\chi_i\,(i=1,2)$ is from the $i$-th $U(1)$ hidden sector with a vector
mass $m_{\chi_i}$, carrying a $U(1)_{i}^{\prime}$ charge $+1$ and is not charged
under SM gauge groups.

In the literature, constraints on the kinetic mixing parameter $\boldsymbol{\delta}$
is usually obtained from a simplified model that the dark photon mixed directly with the photon field.
A more precise treatment is to consider the dark photon mixed with the hypercharge~\cite{Feldman:2007wj,Feng:2023ubl},
where the kinetic mixing parameter is denoted by $\delta$,
as was discussed in this paper.
We translate the experimental constraints on $\boldsymbol{\delta}$ to $\delta$ using the relation~\cite{Feng:2023ubl}
\begin{equation}
	\delta_{1}=\frac{\sqrt{g_{2}^{2}+g_{Y}^{2}}}{g_{2}}\boldsymbol{\delta}\,.
\end{equation}

Now we rewrite the Lagrangian in the gauge eigenbasis $V^{T}=(D,C,B,A^{3})$,
with the kinetic mixing matrix and mass mixing matrix given by
\begin{equation}
\mathcal{K}=\left(\begin{array}{cccc}
1 & \delta_{2} & 0 & 0\\
\delta_{2} & 1 & \delta_{1} & 0\\
0 & \delta_{1} & 1 & 0\\
0 & 0 & 0 & 1
\end{array}\right)\,,\qquad M_{{\rm St}}^{2}=\left(\begin{array}{cccc}
M_{2}^{2}\\
 & M_{1}^{2}\\
 &  & \frac{1}{4}v^{2}g_{Y}^{2} & -\frac{1}{4}v^{2}g_{2}g_{Y}\\
 &  & -\frac{1}{4}v^{2}g_{2}g_{Y} & \frac{1}{4}v^{2}g_{2}^{2}
\end{array}\right)\,,
\end{equation}
which can be diagonalized simultaneously by the rotation matrix $\mathcal{R}$,
transforming the gauge eigenbasis $(D,C,B,A^{3})^{T}$ into mass eigenbasis
$(A^{\prime}_2,A^{\prime}_1,A^\gamma,Z)^{T}$, representing the gauge fields of the two extra dark photons $\gamma^\prime_2, \gamma^\prime_1$, the photon ($\gamma$) and $Z$ boson respectively.
Details of derivation of the rotation matrix $\mathcal{R}$ using perturbation method,
and the full expression of the matrix $\mathcal{R}$
are given in Appendix~\ref{App:2U1dia}.

With the mass eigenbasis, the neutral current couplings are calculated to be
\begin{equation}
\mathcal{L}_{ZA'_{1}A'_{2}A^{\gamma}}=\frac{1}{2}\bar{f_{i}}\gamma^{\mu}\left[\left(v_{i}-a_{i}\gamma^{5}\right)f_{i}Z_{\mu}
+\left(v_{i}^{\prime}-a_{i}^{\prime}\gamma^{5}\right)f_{i}A_{1\mu}^{\prime}
+\left(v_{i}^{\prime\prime}-a_{i}^{\prime\prime}\gamma^{5}\right)f_{i}A_{2\mu}^{\prime}\right]
+eQ_{i}\bar{f_{i}}\gamma^{\mu}f_{i}A_{\mu}^{\gamma}\,,
\end{equation}
where
\begin{align}
v_{i}= & \left(g_{2}\mathcal{R}_{44}-g_{Y}\mathcal{R}_{34}\right)T_{i}^{3}+2g_{Y}\mathcal{R}_{34}Q_{i}\,,\qquad
a_{i}= \left(g_{2}\mathcal{R}_{44}-g_{Y}\mathcal{R}_{34}\right)T_{i}^{3}\,,\\
v_{i}^{\prime}= & \left(g_{2}\mathcal{R}_{42}-g_{Y}\mathcal{R}_{32}\right)T_{i}^{3}+2g_{Y}\mathcal{R}_{32}Q_{i}\,,\qquad
a_{i}^{\prime}=  \left(g_{2}\mathcal{R}_{42}-g_{Y}\mathcal{R}_{32}\right)T_{i}^{3}\,,\\
v_{i}^{\prime\prime}= & \left(g_{2}\mathcal{R}_{41}-g_{Y}\mathcal{R}_{31}\right)T_{i}^{3}+2g_{Y}\mathcal{R}_{31}Q_{i}\,,\qquad
a_{i}^{\prime\prime}=  \left(g_{2}\mathcal{R}_{41}-g_{Y}\mathcal{R}_{31}\right)T_{i}^{3}\,.
\end{align}
In the above equations, $eQ_{i}, T_i$ are respectively the electric charge and the weak isospin of SM fermions.
It's straightforward to obtain the gauge bosons' couplings with charged leptons 
and quarks
\begin{equation}
\mathcal{L}_{\gamma_{1}^{\prime}\bar{f_i}f_i} \approx \left(\delta_{1}g_{Y}c_{W}^{2}Q_i\right)\bar{f_i}\gamma^{\mu}f_{i} A_{1\mu}^{\prime}\,,\qquad
\mathcal{L}_{\gamma_{2}^{\prime}\bar{f_i}f_i} \approx  \left(\delta_{1}\delta_{2}g_{Y}c_{W}^{2}\beta^{2}Q_i \right)\bar{f_i}\gamma^{\mu}f_{i} A_{2\mu}^{\prime}\,,
\end{equation}
where we have defined $\beta^2 \equiv {M_{\gamma_{2}^{\prime}}^{2}}/{M_{\gamma_{1}^{\prime}}^{2}}$.
The couplings of dark photons ($\gamma^\prime_1, \gamma^\prime_2$) to SM neutrinos are however suppressed and can be safely neglected,
which are given by
\begin{align}
\mathcal{L}_{\gamma^{\prime}_1 \bar{\nu}\nu} & \approx-\frac{1}{2}\bigl(\delta_{1}\epsilon^{2}g_{Y}\bigr)\bar{\nu_{L}}\gamma^{\mu}\nu_{L}A_{1\mu}^{\prime}\,,\\
\mathcal{L}_{\gamma^{\prime}_2 \bar{\nu}\nu} & \approx\frac{-\delta_{1}\delta_{2}g_{Y}}{2N_{1}}\left[1+\epsilon^2
(1+\beta^2 )-(1+\epsilon^{2}) (1+ \epsilon^{\prime2})\right]\bar{\nu}_{L}\gamma^{\mu}\nu_{L}A_{2\mu}^{\prime}\nonumber \\
 & \approx\frac{1}{2}\left(\delta_{1}\delta_{2}\epsilon^{2}\epsilon^{\prime2}g_{Y}\right) \bar{\nu}_{L}\gamma^{\mu}\nu_{L} A_{2\mu}^{\prime}\,,
\end{align}
where
\begin{equation}
\epsilon\equiv\frac{M_{1}}{M_{Z}}\approx \frac{M_{\gamma^\prime_1}}{M_{Z}} \,,\qquad
\epsilon^{\prime}\equiv\frac{M_2}{M_{Z}} \approx \frac{M_{\gamma^\prime_2}}{M_{Z}}\,.
\end{equation}

Lastly, the interactions of ($Z,\gamma_{1}^{\prime},\gamma_{2}^{\prime}$)
with dark fermions ($\chi_{1},\chi_{2}$) are given by
\begin{align}
	\mathcal{L}_{\chi_{1}} & =g_{X_{1}}Q_{\chi_{1}}\left(\mathcal{R}_{24}Z_{\mu}+\mathcal{R}_{22}A_{1\mu}^{\prime}+\mathcal{R}_{21}A_{2\mu}^{\prime}\right)\bar{\chi}_{1}\gamma^{\mu}\chi_{1}\,,\\
	\mathcal{L}_{\chi_{2}} & =g_{X_{2}}Q_{\chi_{2}}\left(\mathcal{R}_{14}Z_{\mu}+\mathcal{R}_{12}A_{1\mu}^{\prime}+\mathcal{R}_{11}A_{2\mu}^{\prime}\right)\bar{\chi}_{2}\gamma^{\mu}\chi_{2}\,,
\end{align}
and thus
\begin{align}
	\mathcal{L}_{\gamma_{1}^{\prime}\bar{\chi}_{1}\chi_{1}} & =g_{X_{1}}Q_{\chi_{1}}\bigl(\frac{1}{N_{2}}c_{1}c_{\psi}+\frac{\varepsilon_{12}}{N_{2}}c_{1}s_{2}\bigr)A_{1\mu}^{\prime}\bar{\chi}_{1}\gamma^{\mu}\chi_{1}\nonumber \\
	& \approx g_{X_{1}}Q_{\chi_{1}}A_{1\mu}^{\prime}\bar{\chi}_{1}\gamma^{\mu}\chi_{1}\,,\\
	\mathcal{L}_{\gamma_{2}^{\prime}\bar{\chi}_{1}\chi_{1}} & =g_{X_{1}}Q_{\chi_{1}}\bigl(-\frac{c_{1}s_{2}}{N_{1}}+\frac{\varepsilon_{12}}{N_{1}}c_{1}c_{\psi}+\frac{\varepsilon_{14}}{N_{1}}c_{1}s_{\psi}\bigr)A_{2\mu}^{\prime}\bar{\chi}_{1}\gamma^{\mu}\chi_{1}\nonumber \\
	& \approx g_{X_{1}}Q_{\chi_{1}}\delta_{2}\bigl(\frac{\beta^{2}}{1-\beta^{2}}+\delta_{1}^{2}s_{W}^{2}\bigr)A_{2\mu}^{\prime}\bar{\chi}_{1}\gamma^{\mu}\chi_{1}\,,\\
	\mathcal{L}_{\gamma_{1}^{\prime}\bar{\chi}_{2}\chi_{2}} & =g_{X_{2}}Q_{\chi_{2}}\bigl(-\frac{\varepsilon_{12}}{N_{2}}c_{2}\bigr)A_{1\mu}^{\prime}\bar{\chi}_{2}\gamma^{\mu}\chi_{2}\approx-g_{X_{2}}Q_{\chi_{2}}\bigl(\frac{\delta_{2}}{1-\beta^{2}}\bigr)A_{1\mu}^{\prime}\bar{\chi}_{2}\gamma^{\mu}\chi_{2}\,,\\
	\mathcal{L}_{\gamma_{2}^{\prime}\bar{\chi}_{2}\chi_{2}} & =g_{X_{2}}Q_{\chi_{2}}\bigl(\frac{c_{2}}{N_{1}}\bigr)A_{2\mu}^{\prime}\bar{\chi}_{2}\gamma^{\mu}\chi_{2}\approx g_{X_{2}}Q_{\chi_{2}}A_{2\mu}^{\prime}\bar{\chi}_{2}\gamma^{\mu}\chi_{2}\,.
\end{align}

\subsection{Evolution of the two-step freeze-in darker matter}\label{Sec:Evo2U1}

In this subsection we discuss the full evolution of the two-$U(1)$ model shown in Fig.~\ref{Fig:2U1Model},
using the method developed in~\cite{Aboubrahim:2020lnr,Aboubrahim:2021ycj}.
We assume negligible initial amount of particles in the two $U(1)$ sectors
and $U(1)_{1}^{\prime}$ particles
{$\chi_{1},\gamma_{1}^{\prime}$} are produced through
freeze-in processes from the SM particles: $i\bar{i}\to\chi_{1}\bar{\chi}_{1},\,i\bar{i}\to\gamma_{1}^{\prime}$;
whereas the $U\left(1\right)_{2}^{\prime}$ particles {$\chi_{2},\gamma_{2}^{\prime}$}
are mostly produced from $U(1)_{1}^{\prime}$ sector: $\chi_{1}\bar{\chi}_{1}\to\chi_{2}\bar{\chi}_{2}$, $\chi_{1}\bar{\chi}_{1}\to\gamma_{1}^{\prime}\gamma_{2}^{\prime}$,
$\gamma_{1}^{\prime}\gamma_{1}^{\prime}\to\chi_{2}\bar{\chi}_{2}$
and three-point process $\gamma_{1}^{\prime}\to\chi_{2}\bar{\chi}_{2}$
if $M_{\gamma_{1}^{\prime}}>2m_{\chi_{2}}$.
The freeze-in processes to the $U(1)_{2}^{\prime}$ sector
$\chi_{1}\bar{\chi}_{1}\to\gamma_{2}^{\prime}\gamma_{2}^{\prime}$,
$i\bar{i}\to\chi_{2}\bar{\chi}_{2},\,i\bar{i}\to\gamma_{2}^{\prime}$
can be safely neglected in the computation.
The inverse processes from the $U(1)_{2}^{\prime}$ sector
$\chi_{2}\bar{\chi}_{2}\to\chi_{1}\bar{\chi}_{1}$,
$\chi_{2}\bar{\chi}_{2}\to\gamma_{1}^{\prime}\gamma_{1}^{\prime}$,
$\gamma_{1}^{\prime}\gamma_{2}^{\prime}\to\chi_{1}\bar{\chi}_{1}$
are also disregarded due to their negligible contributions.

The SM, $U(1)_1^\prime$, $U(1)_2^\prime$ sectors, possess different temperatures $T_v, T_{h1}, T_{h2}$ respectively
and evolve almost independently, as a result of the tiny mixings between the three sectors.
These three temperatures are related to each other with functions
 $\eta$($T_{h1}$)$\equiv T_{v}/T_{h1}$
and $\zeta$($T_{h1}$)$\equiv T_{h2}/T_{h1}$,
and can be solved together with the evolution of the hidden sector particles.
The dark particles among each $U(1)$ sector can have rather strong interactions including
$\chi_{1}\bar{\chi}_{1}\leftrightarrow\gamma_{1}^{\prime}\gamma_{1}^{\prime}$, $\chi_{2}\bar{\chi}_{2}\leftrightarrow\gamma_{2}^{\prime}\gamma_{2}^{\prime}$,
and $\chi_{1}\bar{\chi}_{1}\leftrightarrow\gamma_{1}^{\prime}$,
which has significant impacts on the evolution of hidden sector particles.
We assume that $\gamma_{2}^{\prime}$ is the lightest among all hidden sector particles,
ensuring its stability within the age of the Universe.
Such dark photon dark matter can still undergo decay into $e^+e^-$ in minuscule amounts,
contributing to the galactic 511 keV signal,
as we will discussed in Section~\ref{Sec:DP511}.
In the computation, we have taken into account all relevant interactions to calculate the complete evolution of all hidden sector particles.

The evolution of all hidden sector particles within the two-$U(1)$ model
is governed by the
coupled Boltzmann equations for comoving number densities $Y_{\chi_{1}},Y_{\gamma_{1}^{\prime}},Y_{\chi_{2}},Y_{\gamma_{2}^{\prime}}$
and the temperature functions $\eta(T_{h1}),\zeta(T_{h1})$:
\begin{align}
	\frac{\mathrm{d}Y_{\chi_{1}}}{\mathrm{d}T_{h1}}= & -s\frac{\mathrm{d}\rho/\mathrm{d}T_{h1}}{4\rho H}\sum_{i\in\mathrm{SM}}\biggl\{\bigl(Y_{\chi_{1}}^{\mathrm{eq}}\bigr)^{2}\left\langle \sigma v\right\rangle _{\chi_{1}\bar{\chi}_{1}\to i\bar{i}}^{T_{h1}\eta}-Y_{\chi_{1}}^{2}\left\langle \sigma v\right\rangle _{\chi_{1}\bar{\chi}_{1}\to\chi_{2}\bar{\chi}_{2}}^{T_{h1}}\nonumber \\
	& -Y_{\chi_{1}}^{2}\left\langle \sigma v\right\rangle _{\chi_{1}\bar{\chi}_{1}\to\gamma_{1}^{\prime}\gamma_{1}^{\prime}}^{T_{h1}}+Y_{\gamma_{1}^{\prime}}^{2}\left\langle \sigma v\right\rangle _{\gamma_{1}^{\prime}\gamma_{1}^{\prime}\to\chi_{1}\bar{\chi}_{1}}^{T_{h1}}-Y_{\chi_{1}}^{2}\left\langle \sigma v\right\rangle _{\chi_{1}\bar{\chi}_{1}\to\gamma_{1}^{\prime}\gamma_{2}^{\prime}}^{T_{h1}}\nonumber \\
	& +\theta\bigl(M_{\gamma_{1}^{\prime}}-2m_{\chi_{1}}\bigr)\Bigl[-Y_{\chi_{1}}^{2}\left\langle \sigma v\right\rangle _{\chi_{1}\bar{\chi}_{1}\to\gamma_{1}^{\prime}}^{T_{h1}}+\frac{1}{s}Y_{\gamma_{1}^{\prime}}\bigl\langle\Gamma\bigr\rangle_{\gamma_{1}^{\prime}\to\chi_{1}\bar{\chi}_{1}}^{T_{h1}}\Bigr]\biggr\}\,,\label{eq:YchiBol}\\
	\frac{\mathrm{d}Y_{\gamma_{1}^{\prime}}}{\mathrm{d}T_{h1}}= & -s\frac{\mathrm{d}\rho/\mathrm{d}T_{h1}}{4\rho H}\sum_{i\in\mathrm{SM}}\biggl\{ Y_{\chi_{1}}^{2}\left\langle \sigma v\right\rangle _{\chi_{1}\bar{\chi}_{1}\to\gamma_{1}^{\prime}\gamma_{1}^{\prime}}^{T_{h1}}-Y_{\gamma_{1}^{\prime}}^{2}\left\langle \sigma v\right\rangle _{\gamma_{1}^{\prime}\gamma_{1}^{\prime}\to\chi_{1}\bar{\chi}_{1}}^{T_{h1}}+Y_{\chi_{1}}^{2}\left\langle \sigma v\right\rangle _{\chi_{1}\bar{\chi}_{1}\to\gamma_{1}^{\prime}\gamma_{2}^{\prime}}^{T_{h1}}\nonumber \\
	& -Y_{\gamma_{1}^{\prime}}^{2}\left\langle \sigma v\right\rangle _{\gamma_{1}^{\prime}\gamma_{1}^{\prime}\to\chi_{2}\bar{\chi}_{2}}^{T_{h1}}+\theta\bigl(M_{\gamma_{1}^{\prime}}-2m_{i}\bigr)\Bigl[Y_{i}^{2}\left\langle \sigma v\right\rangle _{i\bar{i}\to\gamma_{1}^{\prime}}^{T_{h1}\eta}-\frac{1}{s}Y_{\gamma_{1}^{\prime}}\bigl\langle\Gamma\bigr\rangle_{\gamma_{1}^{\prime}\to i\bar{i}}^{T_{h1}}\Bigr]\nonumber \\
	& +\theta\bigl(M_{\gamma_{1}^{\prime}}-2m_{\chi_{1}}\bigr)\Bigl[Y_{\chi_{1}}^{2}\left\langle \sigma v\right\rangle _{\chi_{1}\bar{\chi}_{1}\to\gamma_{1}^{\prime}}^{T_{h1}}-\frac{1}{s}Y_{\gamma_{1}^{\prime}}\bigl\langle\Gamma\bigr\rangle_{\gamma_{1}^{\prime}\to\chi_{1}\bar{\chi}_{1}}^{T_{h1}}\Bigr]\nonumber \\
	& +\theta\bigl(M_{\gamma_{1}^{\prime}}-2m_{\chi_{2}}\bigr)\Bigl[Y_{\chi_{2}}^{2}\left\langle \sigma v\right\rangle _{\chi_{2}\bar{\chi}_{2}\to\gamma_{1}^{\prime}}^{T_{h1}\zeta}-\frac{1}{s}Y_{\gamma_{1}^{\prime}}\bigl\langle\Gamma\bigr\rangle_{\gamma_{1}^{\prime}\to\chi_{2}\bar{\chi}_{2}}^{T_{h1}}\Bigr]\biggr\}\,,\\
	\frac{\mathrm{d}Y_{\chi_{2}}}{\mathrm{d}T_{h1}}= & -s\frac{\mathrm{d}\rho/\mathrm{d}T_{h1}}{4\rho H}\sum_{i\in\mathrm{SM}}\biggl\{ Y_{\gamma_{1}^{\prime}}^{2}\left\langle \sigma v\right\rangle _{\gamma_{1}^{\prime}\gamma_{1}^{\prime}\to\chi_{2}\bar{\chi}_{2}}^{T_{h1}}+Y_{\chi_{1}}^{2}\left\langle \sigma v\right\rangle _{\chi_{1}\bar{\chi}_{1}\to\chi_{2}\bar{\chi}_{2}}^{T_{h1}}-Y_{\chi_{2}}^{2}\left\langle \sigma v\right\rangle _{\chi_{2}\bar{\chi}_{2}\to\gamma_{2}^{\prime}\gamma_{2}^{\prime}}^{T_{h1}\zeta}\nonumber \\
	& +Y_{\gamma_{2}^{\prime}}^{2}\left\langle \sigma v\right\rangle _{\gamma_{2}^{\prime}\gamma_{2}^{\prime}\to\chi_{2}\bar{\chi}_{2}}^{T_{h1}\zeta}+\theta\bigl(M_{\gamma_{1}^{\prime}}-2m_{\chi_{2}}\bigr)\Bigl[-Y_{\chi_{2}}^{2}\left\langle \sigma v\right\rangle _{\chi_{2}\bar{\chi}_{2}\to\gamma_{1}^{\prime}}^{T_{h1}\zeta}+\frac{1}{s}Y_{\gamma_{1}^{\prime}}\bigl\langle\Gamma\bigr\rangle_{\gamma_{1}^{\prime}\to\chi_{2}\bar{\chi}_{2}}^{T_{h1}}\Bigr]\biggr\}\,,\\
	\frac{\mathrm{d}Y_{\gamma_{2}^{\prime}}}{\mathrm{d}T_{h1}}= & -s\frac{\mathrm{d}\rho/\mathrm{d}T_{h1}}{4\rho H}\sum_{i\in\mathrm{SM}}\Bigl[Y_{\chi_{1}}^{2}\left\langle \sigma v\right\rangle _{\chi_{1}\bar{\chi}_{1}\to\gamma_{1}^{\prime}\gamma_{2}^{\prime}}^{T_{h1}}+Y_{\chi_{2}}^{2}\left\langle \sigma v\right\rangle _{\chi_{2}\bar{\chi}_{2}\to\gamma_{2}^{\prime}\gamma_{2}^{\prime}}^{T_{h1}\zeta}-Y_{\gamma_{2}^{\prime}}^{2}\left\langle \sigma v\right\rangle _{\gamma_{2}^{\prime}\gamma_{2}^{\prime}\to\chi_{2}\bar{\chi}_{2}}^{T_{h1}\zeta}\Bigr]\,,\\
	\frac{\mathrm{d}\eta}{\mathrm{d}T_{h1}}= & -\frac{\eta}{T_{h1}}+\frac{1}{T_{h1}}\left(\frac{4H\rho_{v}+j_{h1}+j_{h2}}{4H\rho_{h1}-j_{h1}}\right)\frac{\mathrm{d}\rho_{h1}/\mathrm{d}T_{h1}}{\mathrm{d}\rho_{v}/\mathrm{d}T_{v}}\,,\\
	\frac{\mathrm{d}\zeta}{\mathrm{d}T_{h1}}= & -\frac{\zeta}{T_{h1}}+\frac{1}{T_{h1}}\left(\frac{4H\rho_{h2}-j_{h2}}{4H\rho_{h1}-j_{h1}}\right)\frac{\mathrm{d}\rho_{h1}/\mathrm{d}T_{h1}}{\mathrm{d}\rho_{h2}/\mathrm{d}T_{h2}}\,,
\label{eq:etaBol}
\end{align}
where the energy transfer densities of two hidden sectors are given
by
\begin{align}
	j_{h1}= & \sum_{i\in\mathrm{SM}}\Bigl[2Y_{i}^{2}s^{2}J_{i\bar{i}\to\chi_{1}\bar{\chi}_{1}}^{T_{h1}\eta}-2Y_{\chi_{1}}^{2}s^{2}J_{\chi_{1}\bar{\chi}_{1}\to\chi_{2}\bar{\chi}_{2}}^{T_{h1}}-Y_{\chi_{1}}^{2}s^{2}J_{\chi_{1}\bar{\chi}_{1}\to\gamma_{1}^{\prime}\gamma_{2}^{\prime}}^{T_{h1}}\nonumber \\
	& -2Y_{\gamma_{1}^{\prime}}^{2}s^{2}J_{\gamma_{1}^{\prime}\gamma_{1}^{\prime}\to\chi_{2}\bar{\chi}_{2}}^{T_{h1}}+\theta\bigl(M_{\gamma_{1}^{\prime}}-2m_{i}\bigr)\Bigl(Y_{i}^{2}s^{2}J_{i\bar{i}\to\gamma_{1}^{\prime}}^{T_{h1}\eta}-Y_{\gamma_{1}^{\prime}}sJ_{\gamma_{1}^{\prime}\to i\bar{i}}^{T_{h1}}\Bigr)\nonumber \\
	& +\theta\bigl(M_{\gamma_{1}^{\prime}}-2m_{\chi_{2}}\bigr)\Bigl(Y_{\chi_{2}}^{2}s^{2}J_{\chi_{2}\bar{\chi}_{2}\to\gamma_{1}^{\prime}}^{T_{h1}\zeta}-Y_{\gamma_{1}^{\prime}}sJ_{\gamma_{1}^{\prime}\to\chi_{2}\bar{\chi}_{2}}^{T_{h1}}\Bigr)\Bigr]\,,\\
	j_{h2}= & \sum_{i\in\mathrm{SM}}\Bigl[2Y_{\chi_{1}}^{2}s^{2}J_{\chi_{1}\bar{\chi}_{1}\to\chi_{2}\bar{\chi}_{2}}^{T_{h1}}+Y_{\chi_{1}}^{2}s^{2}J_{\chi_{1}\bar{\chi}_{1}\to\gamma_{1}^{\prime}\gamma_{2}^{\prime}}^{T_{h1}}+2Y_{\gamma_{1}^{\prime}}^{2}s^{2}J_{\gamma_{1}^{\prime}\gamma_{1}^{\prime}\to\chi_{2}\bar{\chi}_{2}}^{T_{h1}}\nonumber \\
	& +\theta\bigl(M_{\gamma_{1}^{\prime}}-2m_{\chi_{2}}\bigr)\Bigl(-2Y_{\chi_{2}}^{2}s^{2}J_{\chi_{2}\bar{\chi}_{2}\to\gamma_{1}^{\prime}}^{T_{h1}\zeta}+2Y_{\gamma_{1}^{\prime}}sJ_{\gamma_{1}^{\prime}\to\chi_{2}\bar{\chi}_{2}}^{T_{h1}}\Bigr)\Bigr]\,.
\label{eq:BEjh}
\end{align}

In the following we present three benchmark models given in Table~\ref{TableBench_GeV},
with both cases the masses of the dark fermions from the two $U(1)$ sectors are in the GeV range.
In model $x$, all of the dark particles, two dark fermions $\chi_1, \chi_2$, as well as two massive dark photons,
possess the weak scale masses around 100~GeV.
In model $y$, the gauge coupling and the mass of the dark particles from the darker $U(1)_2^\prime$ sector
are chosen such that $\sigma/m \sim 1~{\rm cm^2/g}$ which could potentially explain the cosmic small scale structure problem,
as will be discussed further in Section~\ref{Sec:SIDM}.

\begin{table}
	\begin{center}		
	\resizebox{\textwidth}{8mm}{
\begin{tabular}{|c|c|c|c|c|c|c|c|c|c|c|c|c|}
			\hline
			Model & $m_{\chi_{1}}$ & $m_{\chi_{2}}$ & $M_{\gamma_{1}^{\prime}}$ & $M_{\gamma_{2}^{\prime}}$ & $\delta_{1}$ & $\delta_{2}$ & $g_{X_{1}}$ & $g_{X_{2}}$ & $\Omega_{\chi_{1}}h^{2}$ & $\Omega_{\chi_{2}}h^{2}$ & $\Omega_{\gamma_{2}^{\prime}}h^{2}$ & $\tau_{\gamma_{2}^{\prime}}$/sec \tabularnewline
			\hline
			$x$ & $100$ & $70$ & $160$ & $50$ & $2\times10^{-12}$ & $1\times10^{-10}$ & $0.01$ & $0.03$ & $2.2\times10^{-6}$ & $0.12$ & $3.5\times10^{-9}$ & $2.3\times10^{18}$ \tabularnewline
			\hline
			$y$ & $50$ & $1$ & $70$ & $1.4\times10^{-3}$ & $1.3\times10^{-11}$ & $6.8\times10^{-10}$ & $1\times10^{-3}$ & $0.022$ & $3.6\times10^{-5}$ & $0.12$ & $4.4\times10^{-5}$ & $2.5\times10^{23}$ \tabularnewline
			\hline
			$z$ & $100$ & $30$ & $150$ & $5$ & $8.85\times10^{-12}$ & $3\times10^{-10}$ & $1\times10^{-3}$ & $0.5$ & $1.5\times10^{-6}$ & $8.2\times10^{-3}$ & $0.12$ & $5.1\times10^{17}$ \tabularnewline
			\hline
		\end{tabular}}
	\end{center}
	\caption{Benchmark models with GeV darker matter candidates.
        All masses are in the unit of GeVs.
        The lifetimes of $U(1)_2^\prime$ dark photons $\gamma_2^\prime$ are carefully examined
        across all models to ensure they exceed the age of the Universe.
        Dark fermions $\chi_1, \bar\chi_1$ annihilate efficiently and
        $\gamma_1^\prime$ primarily decays into $\chi_2 \bar\chi_2$, which is the major freeze-in production channel for $\chi_2, \bar\chi_2$.
        For the models $x$ and $y$,
        darker matter $\chi_2, \bar\chi_2$ frozen-in from the dark $U(1)_1^\prime$ sector
        are the primary dark matter candidates with total relic density $0.12$;
        whereas in model $z$,
        the dark photon darker matter $\gamma^\prime_2$ constitute almost the entire dark matter relic density.
        Model $y$ is of particular interest, as the darker matter $\chi_2$, the primary dark matter component of the model,
        exhibits a velocity-dependent annihilation cross-section capable of addressing the cosmic small-scale structure anomalies
        (to be explored in Section~\ref{Sec:SIDM});
        simultaneously, the decay of the dark photon dark matter $\gamma^\prime_2$
        may account for the observed 511~keV photon line signal in the galaxy (to be discussed in Section~\ref{Sec:511}).}
	\label{TableBench_GeV}
\end{table}

\begin{figure}
	\centering
	\subfigure[]{
	\includegraphics[scale=0.5,trim=20 0 30 10,clip]{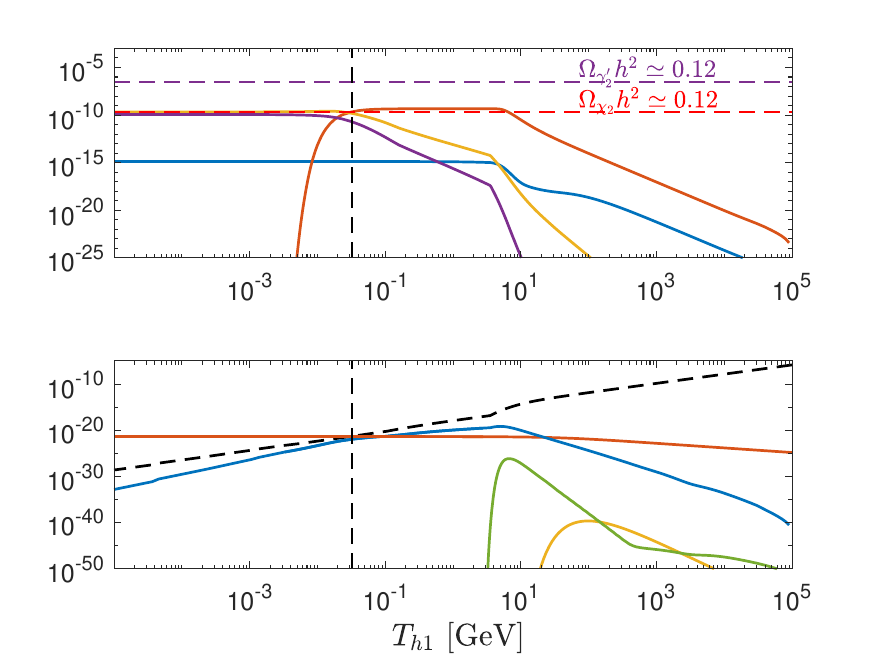} }
	\subfigure[]{
	\includegraphics[scale=0.5,trim=20 0 30 10,clip]{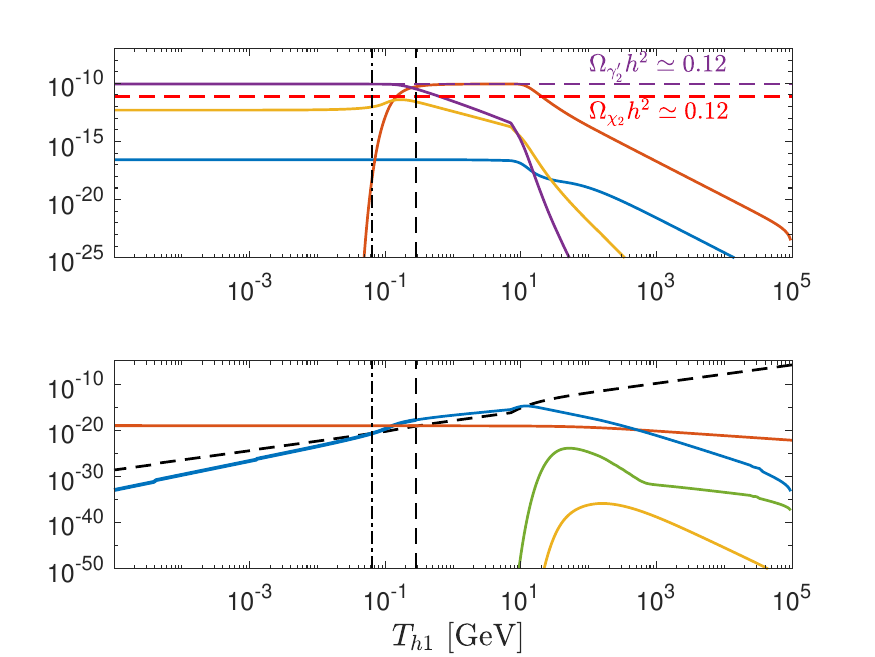} }
	\subfigure{
	\includegraphics[scale=0.25,trim=0 20 0 0,clip]{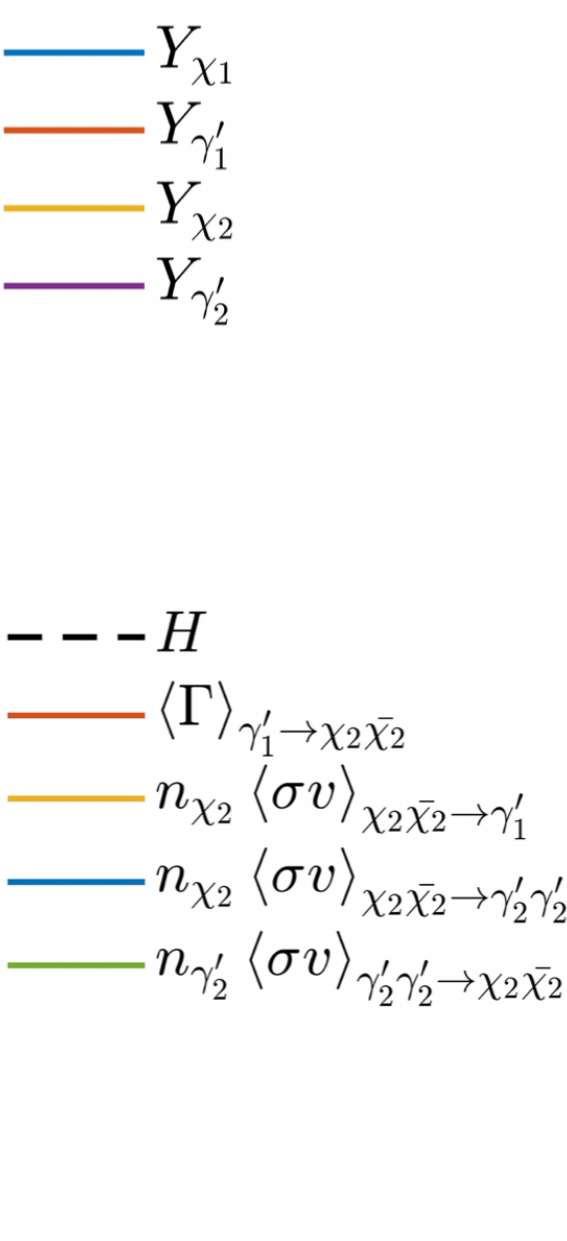} }

		\caption{[Color online] An exhibition of the evolution of
        a dark $U(1)_1$ sector and a darker $U(1)_2$ sector,
        for models $y$ (left) and $z$ (right).
        The upper panels for the two plots show the evolution of the comoving number densities
        of all dark particles in two $U(1)$ hidden sectors,
        and the lower panels of the two plots present the interaction rates of hidden sector interactions compared with the Hubble parameter.
		The red horizontal dash line denotes the observed dark matter relic density for fermion darker matter in the upper panels,
and the purple horizontal dash line denotes the observed dark matter relic density for the dark photon darker matter in the upper panels.
Thus the darker matter $\chi_2$ and $\gamma^\prime_{2}$ originating from the darker hidden sector are respectively
the major component of the dark matter for models $y$ and $z$.
        In the lower panels, interaction rates for the most important interactions within the hidden sectors
        as well as the Hubble expansion rate versus temperature are presented.}
    \label{Fig:YGeV}
\end{figure}

The evolution of dark particles from model $y$ and model $z$ are shown in Fig.~\ref{Fig:YGeV}.
In the lower panel of the two plots,
the vertical black dash line shows the temperature
at which the thermally averaged decay width of $\gamma'_{1}\to\chi_{2}\bar{\chi}_{2}$
goes above the Hubble expansion rate and thus this decay process becomes rapid,
corresponding to a steep drop of the number density of $\gamma^\prime_{1}$ in the upper plot.
The production of dark photon $\gamma^\prime_{2}$ is dominated by the interaction
$\chi_{2}\bar{\chi}_{2}\to\gamma_{2}^{\prime}\gamma_{2}^{\prime}$
within the $U(1)^\prime_{2}$ hidden sector,
although its interaction rate never goes beyond the Hubble expansion rate for model $y$.
Thus the dark photon $\gamma^\prime_{2}$
ultimately constitutes a minuscule fraction of the total dark matter content.
The decay of such dark photon dark matter can account for the 511~keV photon line signal,
as will be discussed in detail in Section~\ref{Sec:511}.

Since the interaction rate of $\chi_2\bar{\chi}_2 \to \gamma^\prime_{2}\gamma^\prime_{2}$
only touches the Hubble line shown in the bottom-left plot,
a clear dark freeze-out of $\chi_2$ is not apparent in this case.
While the dark freeze-out behavior is more evident for model $z$, shown on the right panel.
Here, the moments when the interaction rates of
$\gamma^\prime_{1} \to \chi_2\bar{\chi}_2$ and
$\chi_2\bar{\chi}_2 \to \gamma^\prime_{2}\gamma^\prime_{2}$
drop below the Hubble rate are marked by the black dashed line and black dash-dotted line, respectively.
For model $z$, the darker matter dark photon is the primary dark matter candidate,
originating from the dark freeze-out of $\chi_2$.
In this scenario, the interaction rate of $\chi_{2}\bar{\chi}_{2}\to\gamma_{2}^{\prime}\gamma_{2}^{\prime}$
exceeds the Hubble expansion rate, indicating the rapidity and efficiency of this interaction.
The decay of $\gamma^\prime_{1} \to \chi_2\bar{\chi}_2$ exceeding the Hubble rate
corresponds to a steep drop in the abundance of $\gamma^\prime_{1}$, as shown in the upper plot.
The interaction rate of the dark annihilation process
$\chi_2\bar{\chi}_2 \to \gamma^\prime_{2}\gamma^\prime_{2}$,
denoted by the blue curve, first rises above the Hubble expansion rate,
indicating that thermal equilibrium is established within the $U(1)_2$ hidden sector.
This is the dark freeze-out of $\chi_2$ and
corresponds to a rise in the $\gamma^\prime_2$ number density in the upper plot.
When the interaction rate of $\chi_2\bar{\chi}_2 \to \gamma^\prime_{2}\gamma^\prime_{2}$ drops below the Hubble rate,
as indicated by the black dash-dotted line, the interaction becomes ineffective and the dark freeze-out process concludes.
This corresponds to the abundance of $\gamma^\prime_2$ freezes to its final value in the upper plot.
Simultaneously, the slight decrease in the $\chi_2$ number density also ends, fixing the final relic abundance of $\chi_2$.

\begin{figure}
	\centering
	\subfigure[model $x$]{
		\includegraphics[scale=0.39,trim=23 0 30 0,clip]{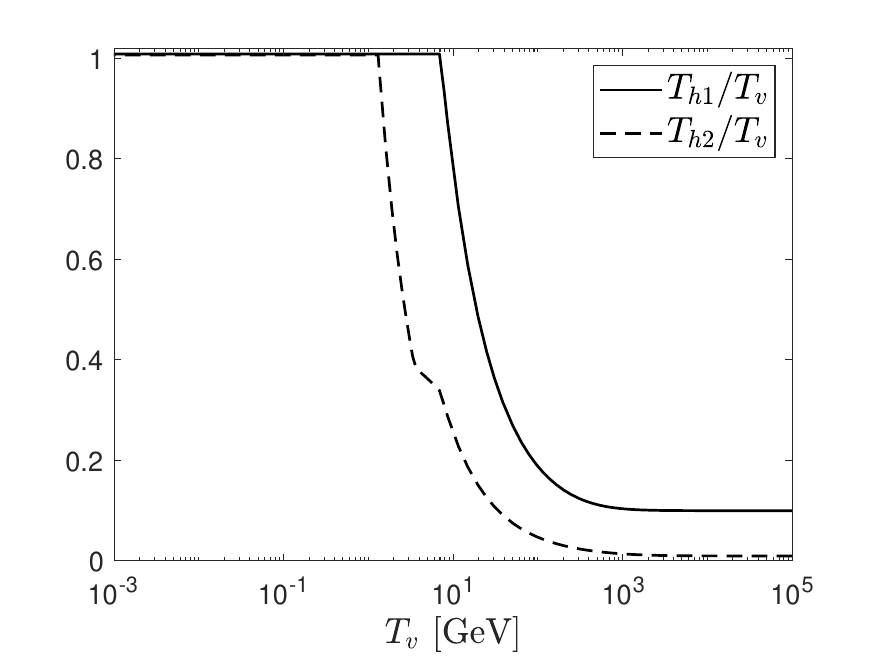} }
	\subfigure[model $y$]{
		\includegraphics[scale=0.39,trim=23 0 30 0,clip]{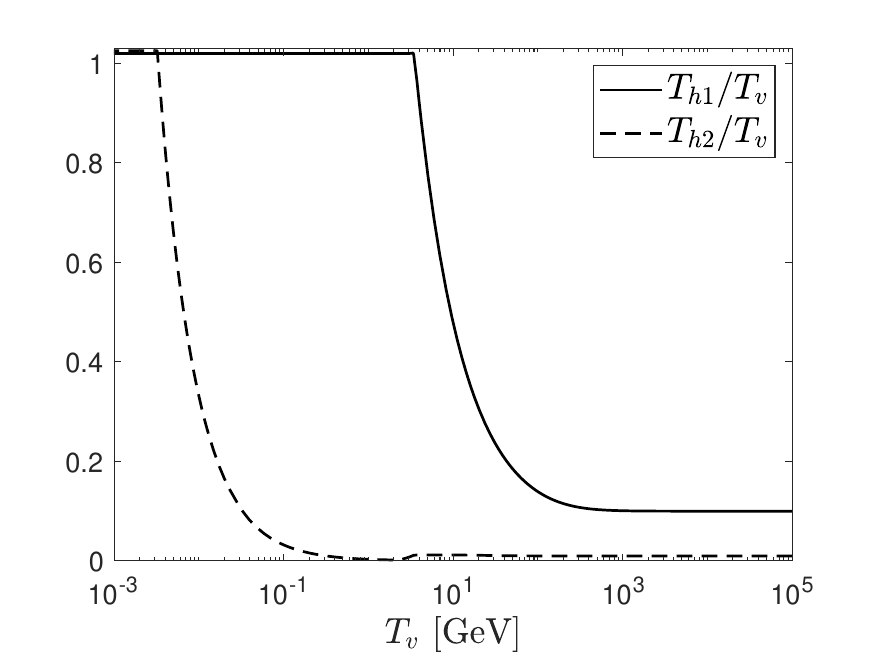} }
	\subfigure[model $z$]{
		\includegraphics[scale=0.39,trim=23 0 30 0,clip]{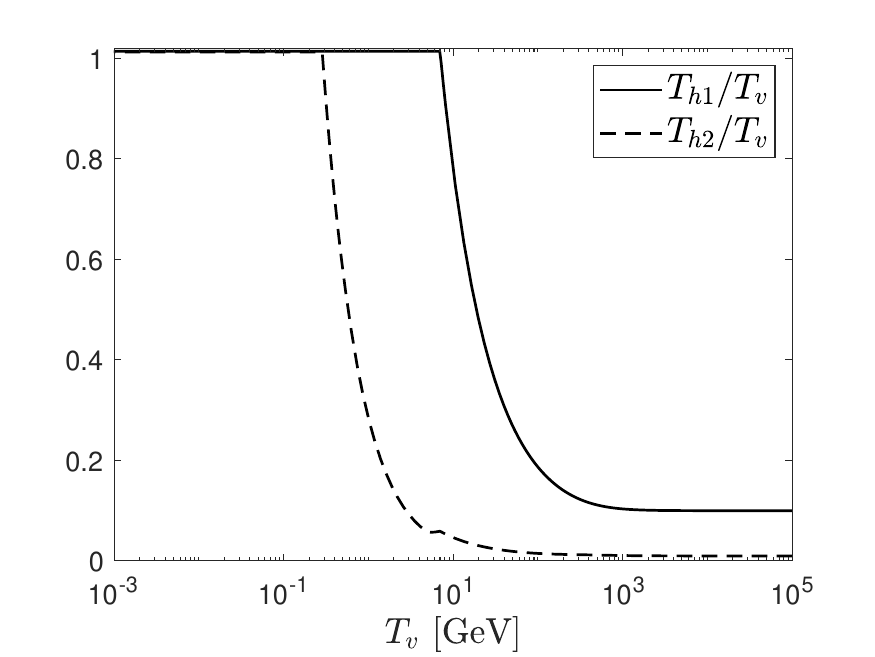} }
	\caption{The hidden sector temperature evolutions of the $U(1)_1$ and $U(1)_2$ sectors for models $x,y,z$ are shown.
    The $U(1)_1$ hidden sector, being directly connected to the SM, thermalized with the visible Universe first;
    while the $U(1)_2$ hidden sector, which is only indirectly coupled to the SM and features a weaker-than-ultraweak connection, thermalized later compared to the $U(1)_1$ sector.}
	\label{Fig:Temp}
\end{figure}

As was discussed in details in~\cite{Feng:2023ubl},
an interaction not achieving equilibrium,
as indicated by its interaction rate never exceeding the Hubble expansion rate,
does not imply that the interaction never took place.
Even ultraweak interactions within the hidden sector can still play a significant role in
determining the final abundance of dark particles in the Universe.

Fig.~\ref{Fig:Temp} shows the temperature evolution of the $U(1)_1$ and $U(1)_2$ hidden sectors for models $x,y,z$,
where the thermalization of both hidden sectors with the visible Universe is clearly observed.

\subsection{SIDM from the darker sector}\label{Sec:SIDM}

SIDM dark matter candidates with velocity-dependent
self-interacting cross-sections
provide a compelling solution to the small scale issues problems observed in the Universe.
SIDM can have rather strong interactions among the dark sector,
and these interaction can be mediated through a dark photon~\cite{Petraki:2014uza,Chen:2015bwa,Duch:2017khv,Duerr:2018mbd,Kamada:2018zxi,Aoki:2018gjf,Kamada:2018kmi,Bernal:2019uqr,Aboubrahim:2020lnr,Du:2020avz,Ghosh:2021wrk}
or other force carriers.

In this subsection, we demonstrate that the darker matter,
which even exhibits weaker-than-ultraweak couplings with the SM sector,
can serve as SIDM candidates responsible for addressing cosmic small-scale structure problems.

The intensity of SIDM dark matter self-interaction is described by~\cite{Tulin:2013teo,Kaplinghat:2015aga}
\begin{equation}
	\frac{\sigma}{m} = \frac{1}{\langle v_r\rangle}\frac{\langle\sigma_T v_r\rangle}{m}\,,
\end{equation}
where the $\sigma_T$ is the energy transfer cross-section
instead of the usual scattering cross-section.\footnote{
The usual scattering cross-section is effectively enhanced in forward-scattering $\cos\theta \to 1$ for light mediator particles.
However, this enhancement does not significantly alter the distribution of dark matter with unchanged trajectories.}
The energy transfer cross-section is given by
\begin{equation}
	\sigma_{T}=\int\frac{{\rm d}\sigma}{{\rm d}\Omega}\bigl(1-\cos\theta\bigr){\rm d}\Omega\,,
\end{equation}
where $\rm{d} \sigma/\rm{d} \Omega$ is the sum of dark matter self-interaction differential cross-sections collected in Appendix \ref{App:SICS},
and the factor $(1-\cos\theta)$ indicates the fractional longitudinal momentum transfer.
Assuming that the dark matter particles possess a Maxwellian velocity distribution,
the velocity-averaged transfer cross-section $\sigma_{T}$ can be derived using
\begin{equation}
	\langle\sigma_{T}v_{r}\rangle=\int_{0}^{v_{r}^{{\rm max}}}\frac{4v_{r}^{2}{\rm e}^{-v_{r}^{2}/v_{0}^{2}}}{\sqrt{\pi}v_{0}^{3}}\sigma_{T}v_{r}{\rm d}v_{r}\,,
\end{equation}
where $v_{r}$
denotes the relative velocity between two interacting DM particles,
and $v_{0}$ is the most probable velocity, related to the mean relative
velocity by $\langle v_{r} \rangle = 2v_{0}/\sqrt{\pi}$~\cite{Kaplinghat:2015aga}.

\begin{figure}
	\begin{center}
		\includegraphics[scale=0.5]{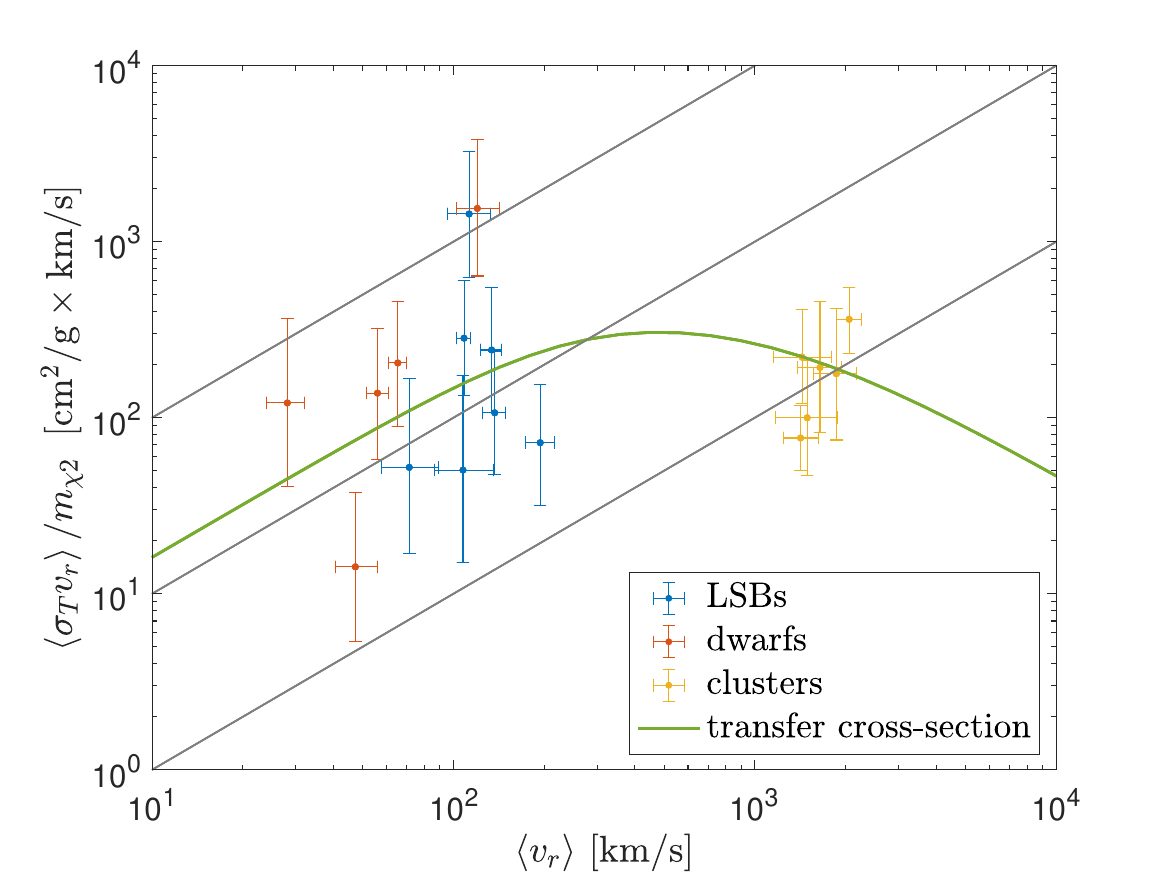}
		\caption{[Color online] An exhibition $\langle\sigma_{T}v_{r}\rangle/m_{\chi}$ for the self-interacting darker matter $\chi_2$
        as a function of the averaged relative velocity $\langle v_{r} \rangle$ for model $y$.
        Data points with error bars from measurements are colored as: LSB galaxies (blue), dwarf galaxies (red), and clusters (yellow).
        The diagonal gray lines correspond to constant values $\sigma/m_{\chi}$ = 0.1, 1 and 10 ${\rm cm^{2}/g}$, respectively.}
	\end{center}
\label{Fig:selfint}
\end{figure}

For model $y$, with its evolution shown in Fig.~\ref{Fig:YGeV}(a),
we plot the
velocity-averaged self-interacting transfer cross-section per unit mass $\langle\sigma_{T}v_{r}\rangle/m_{\chi}$
as a function of the averaged relative velocity $\langle v_{r} \rangle$ in Fig~\ref{Fig:selfint}.
In this case,
the dark matter self-interactions include: $\chi_2 \bar\chi_2 \to \chi_2 \bar\chi_2$,
$\chi_2 \chi_2 \to \chi_2 \chi_2$ and $\bar\chi_2 \bar\chi_2 \to \bar\chi_2 \bar\chi_2$,
mediated by the $U(1)_2$ dark photon $\gamma_2^\prime$.
Data points with error bars are taken from~\cite{Kaplinghat:2015aga},
including low surface brightness galaxies (LSBs) (blue), dwarfs (red) and clusters (yellow).
The velocity-dependent cross-section of the darker matter $\chi_2$ self-interaction in model $y$
effectively fits the observational data from galaxy scales to clusters.

Consequently, the strong self-interactions of the darker matter,
residing in a darker hidden sector and coupling weaker-than-ultraweakly to the SM particles,
could potentially resolve the problems associated with cosmic small-scale structures.

\section{Galactic 511 keV signal from darker hidden sectors}\label{Sec:511}

The galactic 511~keV photon signal has been a longstanding discovery for over half a century,
but without a widely accepted explanation
and now remains as an intriguing but unresolved problem in astrophysics.
In this section, we explore the observed 511 keV signal from the galaxy center
can be attributed to interactions of the darker matter originating from a darker hidden sector.

\subsection{The interpretation of the galactic 511 keV signal}

The galactic 511 keV photon line emission has been firstly observed for more than 50 years~\cite{1st511},
followed by~\cite{511-2,511-3,511-4,511-5,511-6}, and confirmed
by recent measurements including the SPI spectrometer on the INTEGRAL
observatory~\cite{Knodlseder:2003sv,Jean:2003ci,Weidenspointner:2004my,Knodlseder:2005yq,Siegert:2015knp}
and COSI balloon telescope~\cite{Kierans:2019aqz}, see~\cite{Prantzos:2010wi} for an early review.
It was reported in~\cite{Siegert:2015knp} that the total intensity of the galactic
511 keV $\gamma$-ray emission is $(2.74\pm0.25)\times10^{-3}\mathrm{\,ph\,cm^{-2}\,s^{-1}}$
and the bulge intensity is $(0.96\pm0.07)\times10^{-3}\,\mathrm{ph\,cm^{-2}\,s^{-1}}$
at $56\sigma$ significance level.
The distribution of line emission fits to a 2D-Gaussian with full-width-half-maximum (FWHM)
of $\simeq20.55^\circ$ for a broad bulge
and $\simeq5.75^\circ$ for an off-center narrow bulge.
Furthermore, an enhancement of exposure revealed a low surface-brightness disk emission,
resulting in a bulge-to-disk (B/D) flux ratio of $0.58\pm0.13$,
lower than in earlier measurements (B/D$\,\sim 1-3$).

As various astrophysical explanations for the 511 keV signal are inadequate,
exploration into dark matter explanation for this signal has been ongoing for the past 20 years,
encompassing both dark matter annihilation~\cite{Boehm:2003bt,Ascasibar:2005rw,Gunion:2005rw,Huh:2007zw,Vincent:2012an,Wilkinson:2016gsy,Ema:2020fit,Boehm:2020wbt,Drees:2021rsg,DelaTorreLuque:2023cef}
and decay~\cite{Hooper:2004qf,Picciotto:2004rp,Takahashi:2005kp,Finkbeiner:2007kk,Pospelov:2007xh,Cembranos:2008bw,Cai:2020fnq,Vincent:2012an,Lin:2022mqe,DelaTorreLuque:2023cef,Cappiello:2023qwl,Cheng:2023wiz}
to $e^+e^-$ pairs, also see~\cite{Prantzos:2010wi} for an early review.

The 511 keV signal originated from
the annihilation from thermal WIMP
with mass around several MeVs is a well-studied case,
where the light WIMP achieved their final relic abundance through the freeze-out mechanism
with the (total) annihilation cross-section
\begin{equation}
\langle \sigma v \rangle_{\rm ann} \simeq 3\times 10^{-26} \left(\frac{m_{\rm DM}}{{\rm MeV}}\right)^2~{\rm cm^3/s}
\end{equation}
at the freeze-out. While to explain the 511 keV signal the annihilation of dark matter at late times needs to be
\begin{equation}
\langle \sigma v \rangle^{511}_{e^+ e^-} \simeq 5\times 10^{-31} \left(\frac{m_{\rm DM}}{{\rm MeV}}\right)^2~{\rm cm^3/s}\,.\label{eq:CS511}
\end{equation}
The MeV WIMP with an annihilation cross-section featuring a $p$-wave term proportional to the velocity square $\langle \sigma v \rangle = a +bv^2$,
has the potential to explain this phenomenon.
Thus the velocity-squared term needs to be $bv^2 \simeq 3\times 10^{-26}~{\rm cm^3/s}$ at the freeze-out,
and the $s$-wave term $a\simeq 10^{-30}~{\rm cm^3/s}$ is also significant
to compensate the relatively low value (by over one order of magnitude) from $bv^2$ at later times.
MeV scale WIMP dark matter 
will annihilate to either electrons or neutrinos at the freeze-out,
and these processes receive stringent constraints from astrophysics observations,
and are ruled out by the early universe constraints~\cite{Wilkinson:2016gsy}.

Alternative ways of thermally produced dark matter include:
exciting dark matter models~\cite{Finkbeiner:2007kk,Pospelov:2007xh,Cappiello:2023qwl}
the dark matter freeze-out into SM particles other than electrons,
or mostly into extra mediator particles beyond the SM
(such mediators subsequently decay into SM fields).
The dark matter can annihilate to $e^+ e^-$ through
either the loop exchange or through the mediator particle,  and thus explains the 511 keV signal~\cite{Ema:2020fit,Boehm:2020wbt,Drees:2021rsg}.

\subsubsection{Galactic 511 keV signal from dark matter}\label{Sec:DM511}

Low-energy positrons can annihilate with electrons and produce two 511 keV photons directly in a tiny fraction,
or form a bound state known as positronium~\cite{Ps1,Ps2,Ps3} with two possible states:
one is the singlet state (para-positronium/p-Ps) with a zero total spin angular momentum $s=0$,
\begin{equation}
|0,0\rangle =\frac{1}{\sqrt{2}}(\uparrow\downarrow-\downarrow\uparrow)\,,
\end{equation}
which occupies approximately $1/4$ fraction of all the positroniums
and can annihilate into two photons with energies equal to 511 keV;
another is the triplet state (ortho-positronium/o-Ps) with $s=1$,
\begin{equation}
|1,+1\rangle =\uparrow\uparrow\,,\qquad|1,0\rangle =\frac{1}{\sqrt{2}}(\uparrow\downarrow+\downarrow\uparrow)\,,\qquad|1,-1\rangle =\downarrow\downarrow\,,
\end{equation}
which will annihilate into three photons with each energy lower than 511 keV.
Thus the total production rate of 511 keV photons ($\dot{n}_{\gamma}$)
consists of two contributions: (1) positrons annihilate with electrons directly,
(2) annihilation from para-positronium states,
and is given by
\begin{equation}
\dot{n}_{\gamma}=2\Bigl[\bigl(1-f_{p}\bigr)+\frac{1}{4}f_{p}\Bigr]\dot{n}_{e^{+}}=2\Bigl(1-\frac{3}{4}f_{p}\Bigr)\dot{n}_{e^{+}}\,,
\end{equation}
where $\dot{n}_{e^{+}}$ represents the positron production rate,
$f_{p}\approx1$ is the positronium fraction~\cite{Harris:1998tt,Jean:2003ci,Jean:2005af,Beacom:2005qv,Siegert:2015knp}. The total flux of 511 keV $\gamma$-ray can be obtained from integrating
the intensity $I\left(l,b\right)$,  given by an integral
on $\dot{n}_{\gamma}$ along the line of sight (l.o.s)
\begin{equation}
	\Phi_{511}=\int I\left(l,b\right)\mathrm{d}\Omega=\frac{1}{4\pi}\int\mathrm{d}\Omega\int_{\mathrm{l.o.s}}\dot{n}_{\gamma}\left(r\right)\mathrm{d}s\,,\label{eq:flux}
\end{equation}
where $\Omega=2\pi\left(1-\cos\theta\right)$ is the solid angle.
$\theta$ is the angle between the galactic center and the measurement point in the halo,
and $\cos\theta=\cos l\cos b$
where $l$ and $b$ denote the galactic longitude and latitude respectively.

The positron production rates in the case of annihilation and decays
are given by\footnote{
The factor 4 in Eq.~(\ref{eq:rate_ann}) is for a Dirac fermion or a complex scalar.
For a Majorana fermion or a  real scalar ($\bar{X}=X$) this factor should be 2 instead of 4.}
\begin{align}
	\dot{n}_{e^{+}}^{\mathrm{ann}} & =f_{X}\frac{\rho^{2}\left(r\right)}{4m_{X}^{2}}\left \langle \sigma v\right\rangle _{X\bar{X}\to e^{+}e^{-}}\,,\label{eq:rate_ann}\\
	\dot{n}_{e^{+}}^{\mathrm{dec}} & =f_{X}\frac{\rho\left(r\right)}{m_{X}}\Gamma_{X\to e^{+}e^{-}}\mathrm{Br}(X\to e^{+}e^{-})\,,\label{eq:rateDecay}
\end{align}
where $\rho\left(r\right)$ denotes the dark matter density in the
halo of the Milk Way, $\mathrm{Br}(X\to e^{+}e^{-})$ is the branching ratio of $X$ decaying to $e^{+}e^{-}$,
and $f_{X}$ is the fraction of the corresponding
dark matter particle species in the total amount of dark matter.

In this paper, we consider two types of dark matter density profiles widely adopted in the literature:
\begin{enumerate}
	\item The Navarro-Frenk-White (NFW) profile~\cite{Navarro:1995iw}
	\begin{equation}
	\rho_{\mathrm{NFW}}\left(r\right)=\rho_{s}\,{\Bigl(\frac{r}{r_{s}}\Bigr)^{-\gamma}\Bigl(1+\frac{r}{r_{s}}\Bigr)^{\gamma-3}}\,.\label{eq:NFW}
	\end{equation}
	\item The Einasto profile~\cite{Einasto:2009zd}
	\begin{equation}
	\rho_{\mathrm{Einasto}}\left(r\right)=\rho_{s}\exp\biggl\{-\Bigl[\frac{2}{\alpha}\Bigl(\frac{r}{r_{s}}\Bigr)^{\alpha}-1\Bigr]\biggr\}\,.\label{eq:Einasto}
	\end{equation}
\end{enumerate}
Here the parameters $\rho_{s},r_{s},\gamma(\alpha)$,
denoting the scale density, the scale radius, and the slope of halo profiles respectively,
are obtained by fitting the results of N-body simulation.
The galactocentric radius $r$ is given by
\begin{equation}
r=\sqrt{R_{\odot}^{2}+s^{2}-2R_{\odot}s\cos\theta}\,,
\end{equation}
where $s$ is the distance from the Sun to the measurement point in the halo,
$R_{\odot}$ represents the distance from the Sun to the galactic center.

\subsubsection{Constraints of the dark matter interpretation}

\paragraph{Internal bremsstrahlung}

Electromagnetic radiative correction will induce a concurrent
{\it internal bremsstrahlung} process $\chi\bar{\chi}\to e^{+}e^{-}\gamma$
along with the dark matter annihilation~\cite{Beacom:2004pe}.
The diffuse $\gamma$-ray flux from the bremsstrahlung process
at the galactic center must be compatible with the COMPTEL/EGRET data~\cite{Strong:1998ck,Strong:1998fr,Strong:2004de},
and thus constrain the dark matter mass $\lesssim20\mathrm{MeV}$.
This constraint can be relaxed if dark matter is not solely responsible for the 511 keV signal.

\paragraph{Positron in-flight annihilation}

The majority of the positrons will slow down to low energy before annihilating with electrons in the interstellar medium,
while a few energetic positrons may annihilate with electrons during their energy loss (the so-called {\it inflight annihilation}).
The inflight annihilation will emit two photons with energies above 511 keV,
increasing the average diffuse flux in the galactic center.
The corresponding restriction has been derived in~\cite{Beacom:2005qv},
requiring the positron injection energies must be $\lesssim3\mathrm{MeV}$
to match the galactic diffuse $\gamma$-ray data~\cite{Strong:1998ck,Strong:2005zx}.
The inflight annihilation thus imposes constraints,
requiring the annihilating dark matter mass to be $\lesssim3~\mathrm{MeV}$,
to fully explain the 511~keV signal.
For decaying dark matter $X$ to explain the 511 keV signal via the decay $X\to e^+e^-$,
the mass of $X$ must be $\lesssim6~\mathrm{MeV}$.

\paragraph{Additional constraints on feebly interacting particles}\label{Sec:SN-FIPs}

Light feebly interacting particles with masses less than hundreds of MeVs,
could be produced in a supernova core. 
The feeble coupling allows these particles to potentially escape
from the supernova core without interacting with the stellar medium.
The energy loss contributes to the cooling process of supernova,
imposing stringent constraints on the couplings of such particles (e.g., dark photons, sterile neutrinos, etc)
to SM particles~\cite{Chang:2018rso}.
These light feebly interacting particles escaped from the supernova
can undergo further decay to $e^{+}e^{-}$ pairs outside the supernova envelope,
thereby contributing to the 511 keV photon flux.
This imposes additional constraints on the couplings of these particles to SM particles~\cite{Calore:2021klc,Calore:2021lih}.
Especially, additional constraints on sterile neutrinos and dark photons are discussed in~\cite{Calore:2021lih}.
The decay process of sterile neutrinos $\nu_s\to\nu_{\alpha}e^{+}e^{-}$ $(\alpha=e,\mu,\tau)$
can inject positron flux, which contributes to the 511 keV line.
Hence, the detected 511 keV flux implies constraints on the masses of sterile neutrinos and the mixing matrix ($m_{s},|U_{\alpha s}|^{2}$).
Similarly,
a massive dark photon can be produced in the supernova core,
and decay to $e^{+}e^{-}$ outside the supernova envelope,
which generates positron flux.
Thus the 511 keV signal also sets an additional constraint on
the kinetic mixing parameter between the dark photon and the hypercharge (or the photon in effective models),
improving the
bound from SN 1987A energy loss by several orders of magnitude.
We will discuss the application and circumvention of such constraints in our analysis in Section~\ref{Sec:DP511}.

\subsubsection{Annihilating dark matter interpretation}

The 511 keV signal is generally expected to be a consequence of
MeV dark matter particles annihilating to electron-positron pairs.
However, the annihilation of $\mathcal{O}$(MeV) fermionic
dark matter particles annihilate through heavy SM particles (such as the Higgs boson or the $Z$ boson)
will result in an overclosure of the Universe according
to the Lee-Weinberg limit~\cite{Lee:1977ua}.
Later~\cite{Boehm:2003hm} showed that
scalar dark matter candidates with masses lighter than a few GeVs,
where dark matter particles annihilate into SM fermions via exchanging
a charged heavy fermion $F$  through t-channel or a neutral light gauge
boson $Z^{\prime}$ through s-channel, are potentially capable of satisfying both the
dark matter relic density and the observed photon fluxes,
owing to the fact that the annihilation cross-section of spin-0 particles
evades the dependence on dark matter mass.
For example, the annihilation cross-section of a scalar dark matter through a heavy fermion can be roughly written as
\begin{equation}
\sigma_{F-\mathrm{exchange}}\sim\left\{ \begin{array}{c}
{1}/{M_{F}^{2}}\,,\qquad \ \ \ \mathrm{DM\,spin}\ \ 0\\
\\
{m_{\mathrm{DM}}^{2}}/{M_{F}^{4}}\,,\qquad \mathrm{DM\,spin}\,1/2
\end{array}\right.\,
\end{equation}
As more dark matter models have been developed in recent years,
the relic density for MeV dark matter candidates can be satisfied,
offering more possibilities to potentially account for the 511 keV signal.

The viability of light dark matter annihilation models
as the source of the observed 511 keV signal was discussed in~\cite{Boehm:2003bt}.
It was found that in the simplified NFW profile $\rho(r)\propto r^{-\gamma}$ for the inner galactic region,
$\gamma\sim 0.4-0.8$ is preferred for the dark matter interpretation of the 511 keV signal.
It was shown in~\cite{Ascasibar:2005rw},
the thermally averaged cross-section in the velocity-dependent form $\langle \sigma v\rangle =a+bv^{2}$
($a$-term vanishes in the case of $Z^{\prime}$ exchange)
with $a\simeq10^{-31}~\left(m/\mathrm{MeV}\right)^{2}~\mathrm{cm^{3}/s}$
and $bv^{2}\simeq3\times10^{-26}~\mathrm{cm^{3}/s}$ at the freeze-out
satisfies the observed dark matter relic density and 511 $\gamma$-ray flux simultaneously.
The impact of the shape of DM halo profile on 511 keV emission was further explored in~\cite{Ascasibar:2005rw}.
Recently, the sub-GeV scalar dark matter scheme was revisited more comprehensively,
considering exhaustive constraints including laboratory experiments, cosmological and astrophysical observations,
along with its interpretation of the 511 keV signal~\cite{Boehm:2020wbt}.

Other annihilating dark matter models for the 511 keV signal have also been developed in~\cite{Gunion:2005rw,Huh:2007zw,Vincent:2012an,Wilkinson:2016gsy,Ema:2020fit,Drees:2021rsg,DelaTorreLuque:2023cef}. For instance, a $p$-wave annihilation model was constructed in~\cite{Ema:2020fit}
where Majorana fermions dark matter $\chi$ annihilate into $e^{+}e^{-}$ mediated by a real scalar $S$.
A coannihilation model was also constructed in~\cite{Ema:2020fit}
where Majorana fermions $\chi_{1},\chi_{2}$ coannihilate with each other through a dark photon to $e^{+}e^{-}$.
A light Dirac fermion dark matter charged under $U\left(1\right)_{L_{\mu}-L_{\tau}}$ gauge symmetry
which can explain the 511 keV signal was discussed in~\cite{Drees:2021rsg}.

\paragraph{Annihilating freeze-in dark matter explanation is highly implausible}\label{Sec:FDMannN}

Assuming dark matter is initially in equilibrium with SM particles,
its abundance can be reduced through freeze-out annihilations to SM particles
or to additional mediator particles, which will then decay into SM particles.
Both direct and indirect detections impose stringent constraints on freeze-out dark matter models,
making it more challenging to interpret the 511 keV signal using thermal WIMPs.
Freeze-in mechanism is a novel mechanism for the dark matter generation~\cite{Hall:2009bx,Chu:2011be,Bernal:2017kxu},
which can produce the observed value of dark matter relic density
while also circumvent constraints from direct and indirect detections.

However, it is highly implausible for freeze-in dark matter $\chi$ to be responsible for the 511 keV signal through annihilation.
If the 511 keV signal is attributing to the annihilation $\chi \bar{\chi} \to e^+ e^-$,
the effective coupling between $\chi\bar\chi$ and $e^+ e^-$ will be too large for the freeze-in production,
which will result in an overabundance of $\chi$ production.
If the hidden sector possesses more than one particle species,
e.g., a scalar or a dark photon, which will eventually decay to $e^+ e^-$ before BBN,
one can arrange the hidden sector interaction in such a way
that overproduced $\chi$ annihilates into these mediator particles.
However, in this scenario, the mediator will encounter stringent constraints
from astrophysical observations, especially from supernova~\cite{Chang:2018rso,Calore:2021klc,Calore:2021lih},
rendering them already ruled out by experiments.

\subsubsection{Decaying dark matter interpretation}

As the source of the galactic 511 keV signal remains unconfirmed,
the decaying dark matter explanation which was actively researched,
is also one of the significant candidates of the 511 keV signal~\cite{Hooper:2004qf,Picciotto:2004rp,Takahashi:2005kp,Finkbeiner:2007kk,Pospelov:2007xh,Cembranos:2008bw,Cai:2020fnq,Lin:2022mqe,DelaTorreLuque:2023cef,Cappiello:2023qwl,Cheng:2023wiz}.
Dark matter with lifetime greater than the age of the Universe,
decays into $e^{+}e^{-}$ pairs could potentially account for the observed signal.
Though it has been suggested the emission originates from dark matter decay face challenges,
as they seem inconsistent with the observed data,
particularly due to its comparatively broad distribution~\cite{Vincent:2012an,Ascasibar:2005rw}.

However, given the uncertainties inherent in astrophysical measurements,
the decaying dark matter explanation continues to be one of several competing theories requiring further investigation.
Furthermore, we cannot dismiss the possibility that dark matter density surge towards the galactic center
resulting from adiabatic compression under the influence of massive black holes~\cite{Prada:2004pi}
as well as other gravitational effects,
leading to a steeper profile of the dark matter halo,
which provide much better fit for the observed morphology.

Specifically, it is shown in~\cite{Picciotto:2004rp} that
the decay of unstable relic particles with $\mathcal{O}$(MeV) masses can be
can be a source of the 511 keV line.
Although a considerable part of the parameter space for sterile neutrinos with masses ranging from 1 to 50 MeV
is excluded due to the overabundance of positron flux.
Light decaying axinos in an R-parity violating supersymmetric model were explored in~\cite{Hooper:2004qf},
finding that a cusped halo profile is in accordance with the INTEGRAL measurements.
It is shown that $\gamma\sim 0.8-1.5$ in the simplified NFW profile $\rho(r)\propto r^{-\gamma}$ for the inner galactic region,
steeper than that in the annihilation dark matter model.

As an alternative, the exciting dark matter model was studied in~\cite{Finkbeiner:2007kk,Pospelov:2007xh,Cappiello:2023qwl},
where excited heavy dark matter $\chi^{\ast}$ is emanated from inelastic
scattering $\chi\chi\to\chi\chi^{\ast}$ in the galactic center and subsequently transits back to the ground state through
$\chi^{\ast}\to\chi e^{+}e^{-}$ if $m_{\chi^{\ast}}-m_{\chi}\gtrsim2m_{e}$,
emitting the expected 511 keV gamma ray signal.
For this type of models, the positron injection energy only constrains the
mass splitting between the excited and the ground states, rather than the dark matter mass.
The excitation rate decreases significantly with galactocentric radius,
resulting in a radial cutoff and a less cuspy signal compared to annihilation dark matter.

\subsection{511 keV signal from dark photon darker matter}\label{Sec:DP511}

In this subsection,
we explore the possibility that the 511 keV signal attributed to the dark photon darker matter generated through a two-step freeze-in process.
Decaying dark matter with an extremely tiny coupling $g$ with the $e^+e^-$ pair
is capable of explaining the galactic 511 keV signal, and based on dark matter profiles widely adopted, one needs
\begin{equation}
g^2 \times f_X \times {\rm Br}(X\to e^+e^-) \sim 10^{-50} - 10^{-47}\,,
\end{equation}
where $f_X$ is the fraction of $X$ in the total dark matter relic density,
and ${\rm Br}(X\to e^+e^-)$ is the branching fraction of $X$ decay to $e^+e^- $.
This condition is so restrictive that the decaying dark matter $X$ cannot be produced from the usual freeze-in.
Especially, the coupling $g\lesssim 10^{-18}$ is too tiny even for the freeze-in production of dark matter.

The two-$U(1)$ model as discussed in the previous section,
featuring a two-step freeze-in generation of the darker matter,
can provide a dark photon darker matter with ultraweak coupling strength with the SM sector.
Such dark photon darker matter, even with a lifetime exceeding the age of the Universe,
can still decay in minuscule amounts into $e^+e^-$ pairs,
potentially contributing to the 511 keV signal.
The tiny kinetic mixing $\delta_1\lesssim 10^{-10}$ can generate
$U(1)_{1}^{\prime}$ sector particles via freeze-in mechanism,
which can further produce $U(1)_{2}^{\prime}$ sector particles through freeze-in.
The coupling between $U(1)_{2}^{\prime}$ and the SM is through the $U(1)_{1}^{\prime}$ sector
and thus the coupling of $U(1)_{2}^{\prime}$ to the SM fermions is indirect
and is proportional to $\delta_1\times \delta_2 \lesssim 10^{-18}$,
which can potentially explain the 511 keV signal.
The dark photon $\gamma_{1}^{\prime}$ will decay to $e^{+}e^{-}$ before BBN.
Dark matter candidates are generally a combination of $\chi_{1},\chi_{2},\gamma_{2}^{\prime}$,
among which $\gamma_{2}^{\prime}$ with mass $2m_{e}\lesssim M_{\gamma_{2}^{\prime}}\lesssim6\mathrm{MeV}$,
only occupies a tiny fraction of the total dark matter relic density
and is responsible for the 511~keV signal through the decay to $e^{+}e^{-}$ pairs.
In the benchmark models we analyzed,
the mass of $\gamma_2^\prime$ is set to 6~MeV to explain the 511 keV signal through the decaying process
$\gamma_2^\prime \to e^+e^-$,
while the mass of $\gamma_1^\prime$ can range from tens of MeV to hundreds of GeVs.

\begin{figure}
	\centering
	\subfigure[]{
    \includegraphics[scale=0.32]{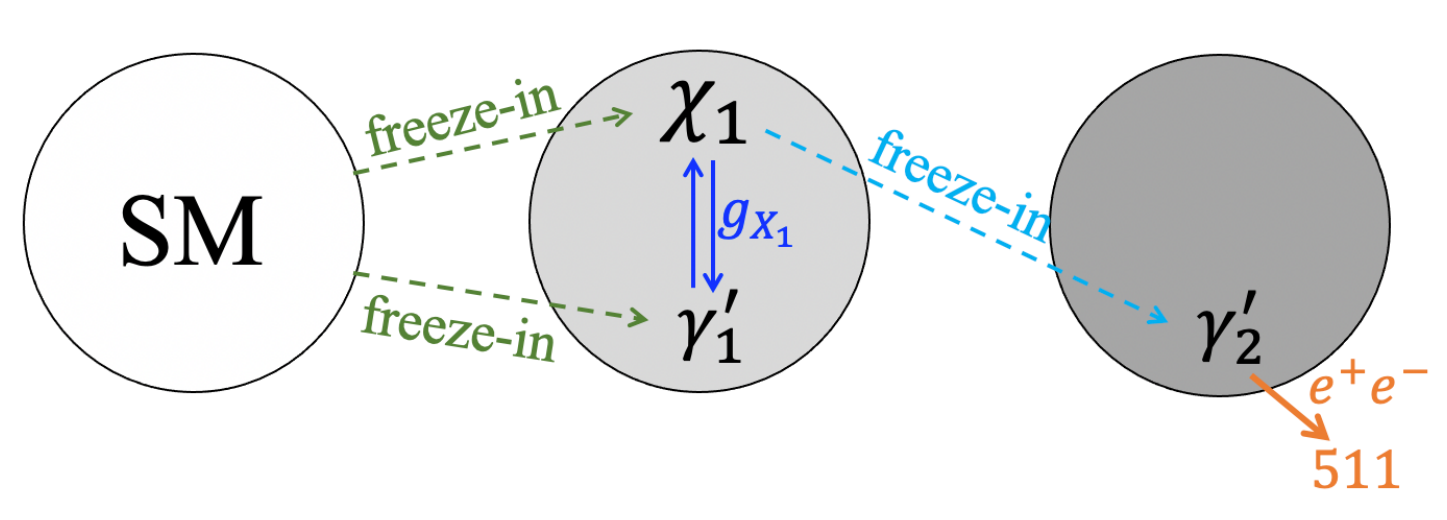} }
	\subfigure[]{
	\includegraphics[scale=0.32]{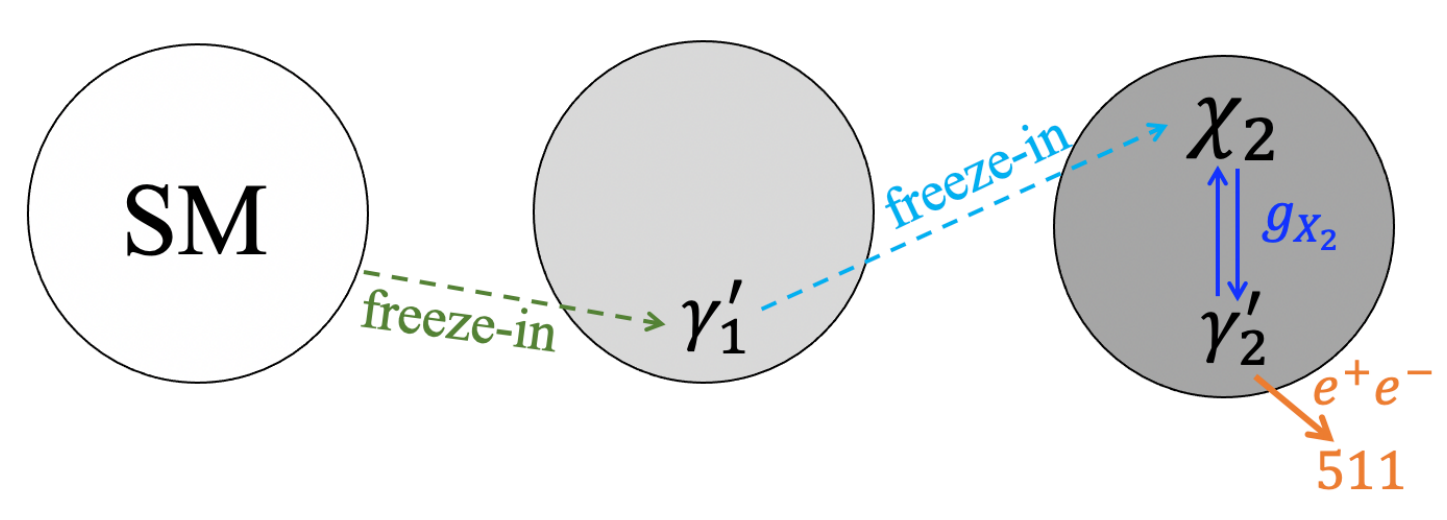} }
    \subfigure[]{
	\includegraphics[scale=0.6,trim=20 0 30 0,clip]{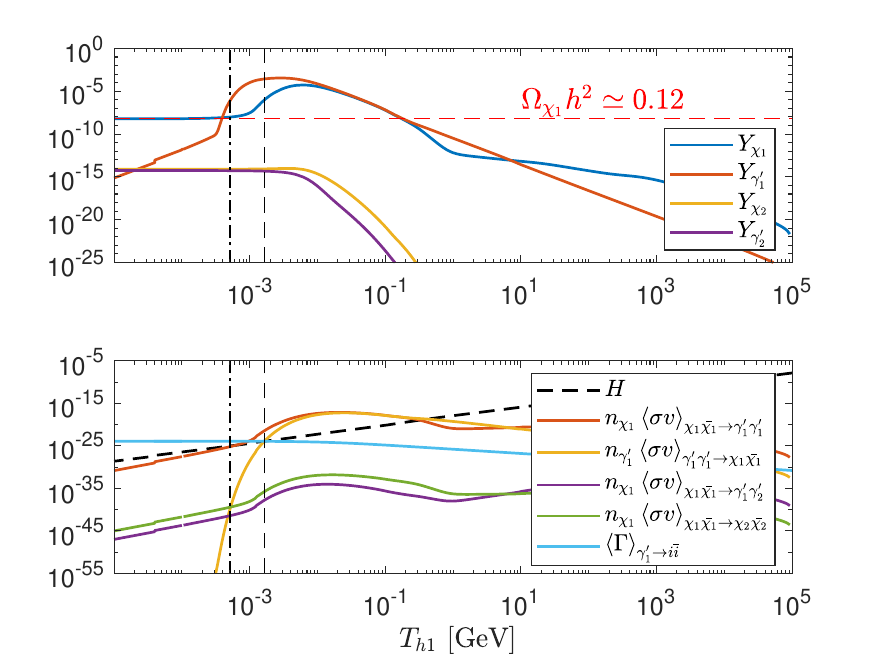} }
    \subfigure[]{
	\includegraphics[scale=0.6,trim=20 0 30 0,clip]{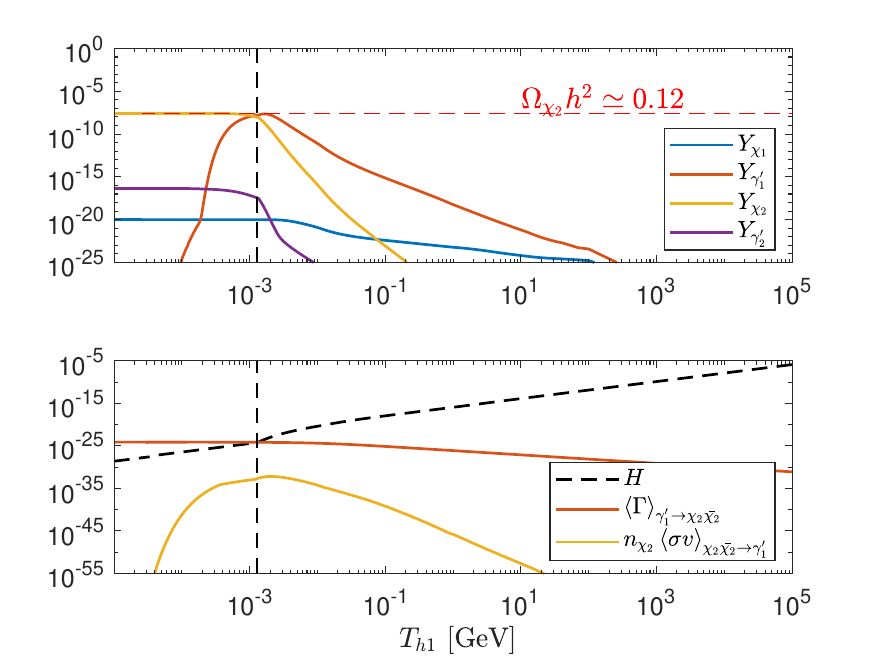} }
	\caption{ \label{Fig:Evo} [Color online] An exhibition of the evolution for the two $U(1)$ hidden sectors,
        for Case 1 (using model~$a$ as an illustration) and Case 2 (using model~$d$ as an illustration).
        In Case~1, we arrange the parameters in such a way that
        $\chi_1$ is the main component of dark matter;
        whereas in Case~2, $\chi_2$ is the main component of dark matter.
        Thus the two-$U(1)$ model can be reduced to Panel (a) for Case 1 and Panel (b) for Case 2.
        Panels (c) and (d) show the evolution of the hidden sector particles
        as well as the interaction rates of hidden sector interactions compared with the Hubble parameter,
        for model $a$ and model $d$ respectively.
		In Panels (c) and (d), the above figure shows the evolution of the comoving number densities
        of all dark particles in two $U(1)$ hidden sectors.
		The red horizontal line denoting the observed dark matter relic density.
        In the lower figures, interaction rates for various interactions among the hidden sectors
        as well as the Hubble parameter versus temperature are presented.}
\end{figure}

In certain parameter space
one can easily arrange either $\chi_1$ or $\chi_2$ constitutes almost the entire dark matter relic density,
and thus the model shown in Fig.~\ref{Fig:2U1Model} can be
simplified to two models shown in Fig.~\ref{Fig:Evo}(a) and \ref{Fig:Evo}(b).
Thus we explore two cases to explain the 511 keV flux, each possessing
three components of the dark matter candidates $\{\chi_{1},\chi_{2},\gamma_{2}^{\prime}\}$
offered by both two $U(1)$ sectors with different proportion
in the total dark matter relic density.
For each case, we consider three different benchmark models depending on the masses of $\gamma'_{1}$
and $\chi_{1}$: $M_{\gamma'_{1}}<m_{\chi_{1}}$, $m_{\chi_{1}}<M_{\gamma'_{1}}<2m_{\chi_{1}}$,
and $M_{\gamma'_{1}}>2m_{\chi_{1}}$.
\begin{description}
	\item [{Case}] \textbf{1:} {\bf $M_{\gamma'_{1}}<2m_{\chi_{2}}$ (the dark matter is predominantly $\chi_{1}$)}
     In this case, the $U(1)^\prime_{2}$ sector particles
	$\chi_{2},\gamma'_{2}$ are produced via freeze-in from the $U(1)^\prime_{1}$ sector
	through four-point interactions $\gamma_{1}^{\prime}\gamma_{1}^{\prime}\to\chi_{2}\bar{\chi}_{2}$,
    $\chi_{1}\bar{\chi}_{1}\to\chi_{2}\bar{\chi}_{2}$,
    $\chi_{1}\bar{\chi}_{1}\to\gamma_{1}^{\prime}\gamma_{2}^{\prime}$.
	Particles within each hidden sector can have vigorous interactions
	including $\chi_{1}\bar{\chi}_{1}\leftrightarrow\gamma_{1}^{\prime}$ or
    $\chi_{1}\bar{\chi}_{1}\leftrightarrow\gamma_{1}^{\prime}\gamma_{1}^{\prime}$,
	$\chi_{2}\bar{\chi}_{2}\leftrightarrow\gamma_{2}^{\prime}\gamma_{2}^{\prime}$.
	As the temperature drops down, the $U(1)^\prime_{1}$ hidden sector will
	encounter the freeze-out of $\chi_{1}$ or $\gamma'_{1}$ depending
	on their masses. Eventually all of the dark photon $\gamma_{1}^{\prime}$ will decay
	into SM fermions through $\gamma_{1}^{\prime}\to i\bar{i}$ before BBN,
    and the remaining $\chi_{1}$ constitutes the dominant dark matter candidate.
    The	dark particles $\chi_{2}, \gamma_{2}^{\prime}$ occupy only a tiny fraction of
	the total dark matter relic density.
    The 511 keV signal is interpreted by the decay $\gamma_{2}^{\prime}\to e^{+}e^{-}$.
	\item [{Case}] \textbf{2: } {\bf $M_{\gamma_{1}^{\prime}}>2m_{\chi_{2}}$ (the dark matter is predominantly $\chi_{2}$)}
    In this case, the dark fermion $\chi_{2}$
	is mainly produced by the intensive decay process $\gamma'_{1}\to\chi_{2}\bar{\chi}_{2}$,
	and can interact with the dark photon $\gamma_{2}^{\prime}$ through
    $\chi_{2}\bar{\chi}_{2}\leftrightarrow\gamma_{2}^{\prime}\gamma_{2}^{\prime}$,
	producing the desired amount of the dark photon $\gamma'_{2}$ to explain the 511 keV signal.
    The $U(1)^\prime_{1}$ gauge boson $\gamma_{1}^{\prime}$
	can also interchange with $\chi_{1}$ or decay into SM fermions and finally reach a negligible amount.
    The dark matter can primarily consist of $\chi_{2}$,
    with minor contribution from $\chi_{1}$ and $\gamma'_{2}$.
	The 511 keV signal is also attributed to the dark photon $\gamma'_{2}$ via its decay into $e^{+}e^{-}$.
\end{description}

The model shown in Fig.~\ref{Fig:2U1Model} is robust.
In addition to the benchmark models shown in Table~\ref{TableBench},
there is plenty of parameter space available for adjustment,
such that both $\chi_1$ and $\chi_2$ constitute significant fractions of the total dark matter relic density.

\begin{figure}
	\begin{center}
			\includegraphics[scale=0.74]{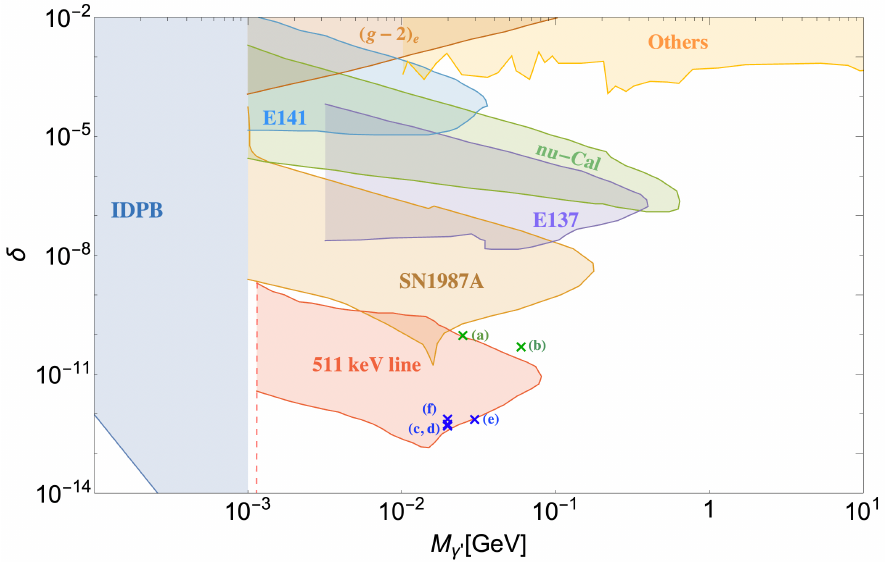}
	    	\caption{[color online] A display of current experimental constraints (colored regions) on the kinetic mixing parameter.
            We also include the constraint from 511 keV signal discussed in Section~\ref{Sec:SN-FIPs} (the region in red)~\cite{Calore:2021lih}.
            This constraint is valid for a dark photon that can decay only into $e^+e^-$.
            Benchmark models $a$ and $b$ fall into this category, thus we choose parameters outside this red region.
            In benchmark models $c$-$g$, the dark photon $\gamma_1^\prime$ can also decay into dark matter,
            allowing this constraint to be circumvented.
            Hence, we can safely choose parameter values within the red region.}
			\label{Fig:Bound}
	 \end{center}
\end{figure}

As discussed in Section~\ref{Sec:SN-FIPs}, $\mathcal{O}$(MeV) scale particles
interacting feebly with SM fields
can be produced in a supernova core
and decay into electron-positron pairs outside the supernova envelope,
subject to stringent constraints from the observed 511 keV photon emission.
The energy loss of SN 1987A as well as various laboratory experiments
also impose limits on the model parameters.
The current experimental constraints on the kinetic mixing parameter $\delta_1$
in terms of the dark photon mass is plotted in Fig.~\ref{Fig:Bound} with all
our benchmark models marked respectively.

\begin{table}
	\centering
	\resizebox{\textwidth}{16mm}{
		\begin{tabular}{|c|c|c|c|c|c|c|c|c|c|c|c|c|}
			\hline
			\multicolumn{1}{|c|}{Case} &
			\multicolumn{1}{c|}{Model} &
			\multicolumn{1}{c|}{$m_{\chi_{1}}$} &
			\multicolumn{1}{c|}{$m_{\chi_{2}}$} &
			\multicolumn{1}{c|}{$M_{\gamma_{1}^{\prime}}$} &
			\multicolumn{1}{c|}{$\delta_{1}$} &
			\multicolumn{1}{c|}{$\delta_{2}$} &
			\multicolumn{1}{c|}{$g_{X_{1}}$} &
			\multicolumn{1}{c|}{$g_{X_{2}}$} &
			\multicolumn{1}{c|}{$\Omega_{\chi_{1}}h^{2}$} &
			\multicolumn{1}{c|}{$\Omega_{\chi_{2}}h^{2}$} &
			\multicolumn{1}{c|}{$\Omega_{\gamma_{2}^{\prime}}h^{2}$} &
			\multicolumn{1}{c|}{$\tau_{\gamma_{2}^{\prime}}$} \\ \hline

			\multicolumn{1}{|c|}{\multirow{3}{*}{$M_{\gamma_{1}^{\prime}}<2m_{\chi_{2}}$}} &
			\cellcolor{green!10}{$a$} &
			\cellcolor{green!10}{$35$}&
			\cellcolor{green!10}{$30$}&
			\cellcolor{green!10}{$25$}&
			\cellcolor{green!10}{$1\times10^{-10}$} &
			\cellcolor{green!10}{$9.5\times10^{-9}$} &
			\cellcolor{green!10}{$6.15\times10^{-3}$}  &
			\cellcolor{green!10}{$0.08$}&
			\cellcolor{green!10}{$0.12$}&
			\cellcolor{green!10}{$10^{-7}$}&
			\cellcolor{green!10}{$6.5\times10^{-9}$}&
			\cellcolor{green!10}{$1.6\times10^{19}$ }  \\ \cline{2-13}
			
			\multicolumn{1}{|c|}{} &
			\cellcolor{green!10}{$b$} &
			\cellcolor{green!10}{$40$}&
			\cellcolor{green!10}{$35$}&
			\cellcolor{green!10}{$60$}&
			\cellcolor{green!10}{$5.2\times10^{-11}$} &
			\cellcolor{green!10}{$4.5\times10^{-7}$} &
			\cellcolor{green!10}{$3.8\times10^{-5}$}  &
			\cellcolor{green!10}{$0.92$}&
			\cellcolor{green!10}{$0.12$}&
			\cellcolor{green!10}{$10^{-10}$}&
			\cellcolor{green!10}{$6.8\times10^{-10}$}&
			\cellcolor{green!10}{$9.8\times10^{17}$ }  \\ \cline{2-13}
			
			\multicolumn{1}{|c|}{} &
			\cellcolor{blue!10}{$c$} &
			\cellcolor{blue!10}{$8$}&
			\cellcolor{blue!10}{$25$}&
			\cellcolor{blue!10}{$20$}&
			\cellcolor{blue!10}{$5.8\times10^{-13}$} &
			\cellcolor{blue!10}{$5\times10^{-6}$} &
			\cellcolor{blue!10}{$7\times10^{-3}$}  &
			\cellcolor{blue!10}{$0.5$}&
			\cellcolor{blue!10}{$0.12$}&
			\cellcolor{blue!10}{$10^{-12}$}&
			\cellcolor{blue!10}{$2.8\times10^{-10}$}&
			\cellcolor{blue!10}{$6.6\times10^{17}$ }  \\ \hline
			
			\multicolumn{1}{|c|}{\multirow{4}{*}{$M_{\gamma_{1}^{\prime}}>2m_{\chi_{2}}$}} &
			\cellcolor{blue!10}{$d$} & 
			\cellcolor{blue!10}{$30$}&
			\cellcolor{blue!10}{$9$}&
			\cellcolor{blue!10}{$20$}&
			\cellcolor{blue!10}{$5.3\times10^{-13}$} &
			\cellcolor{blue!10}{$3\times10^{-6}$} &
			\cellcolor{blue!10}{$1\times10^{-5}$}  &
			\cellcolor{blue!10}{$2.5\times10^{-5}$}&
			\cellcolor{blue!10}{$1.7\times10^{-13}$}&
			\cellcolor{blue!10}{$0.12$}&
			\cellcolor{blue!10}{$9.4\times10^{-10}$}&
			\cellcolor{blue!10}{$2.2\times10^{18}$ } \\ \cline{2-13}
			
			\multicolumn{1}{|c|}{} &
			\cellcolor{blue!10}{$e$} &
			\cellcolor{blue!10}{$20$}&
			\cellcolor{blue!10}{$14$}&
			\cellcolor{blue!10}{$30$}&
			\cellcolor{blue!10}{$7.6\times10^{-13}$} &
			\cellcolor{blue!10}{$4\times10^{-6}$} &
			\cellcolor{blue!10}{$1\times10^{-5}$}  &
			\cellcolor{blue!10}{$3.7\times10^{-5}$}&
			\cellcolor{blue!10}{$5.0\times10^{-11}$}&
			\cellcolor{blue!10}{$0.12$}&
			\cellcolor{blue!10}{$2.0\times10^{-9}$}&
			\cellcolor{blue!10}{$3.4\times10^{18}$ } \\ \cline{2-13}
			
			\multicolumn{1}{|c|}{} &
			\cellcolor{blue!10}{$f$} &
			\cellcolor{blue!10}{$8$}&
			\cellcolor{blue!10}{$9$}&
			\cellcolor{blue!10}{$20$}&
			\cellcolor{blue!10}{$7.8\times10^{-13}$} &
			\cellcolor{blue!10}{$4.5\times10^{-6}$} &
			\cellcolor{blue!10}{$1\times10^{-12}$}  &
			\cellcolor{blue!10}{$2\times10^{-5}$}&
			\cellcolor{blue!10}{$1.3\times10^{-5}$}&
			\cellcolor{blue!10}{$0.12$}&
			\cellcolor{blue!10}{$3.8\times10^{-10}$}&
			\cellcolor{blue!10}{$4.9\times10^{17}$ }  \\ \cline{2-13}
			
			\multicolumn{1}{|c|}{} &
			\cellcolor{blue!10}{$g$} &
			\cellcolor{blue!10}{$10^{5}$}&
			\cellcolor{blue!10}{$8\times10^{4}$}&
			\cellcolor{blue!10}{$2\times10^{5}$}&
			\cellcolor{blue!10}{$5.8\times10^{-12}$} &
			\cellcolor{blue!10}{$1\times10^{-10}$} &
			\cellcolor{blue!10}{$1\times10^{-3}$}  &
			\cellcolor{blue!10}{$0.6$}&
			\cellcolor{blue!10}{$2.6\times10^{-7}$}&
			\cellcolor{blue!10}{$0.12$}&
			\cellcolor{blue!10}{$4.8\times10^{-5}$}&
			\cellcolor{blue!10}{$1.3\times10^{23}$ }  \\ \hline

	\end{tabular}}
\caption{The benchmark models we consider in this work for Case~1 and Case~2.
All masses are in the unit of MeVs.
In both cases, $M_{\gamma_{2}^{\prime}}$ is taken to be 6~MeV to explain the 511~keV signal.
The lifetimes of the two dark photons (in the unit of seconds) for each model are listed in the last two columns,
ensuring $\gamma^\prime_1$ decays before BBN while the lifetime of $\gamma^\prime_2$  exceeds the age of the Universe.
All benchmark models satisfy all relevant experimental constraints and are also marked in Fig.~\ref{Fig:Bound}.
In both of Case~1 and Case~2,
the parameter values can be adjusted
such that both $\chi_1$ and $\chi_2$ constitute significant fractions of the total dark matter relic density.}
\label{TableBench}
\end{table}

As shown in Fig.~\ref{Fig:Bound},
in addition to the SN 1987A bound on the extra $U(1)$,
the 511~keV signal provide more stringent constraints on the dark photon mixing~\cite{Calore:2021lih},
which is the most sensitive region for our analysis.
The 511~keV signal constraint is valid for models that the dark photon can decay only into $e^+e^-$,
In models $a$ and $b$, marked by the shaded light green in Table~\ref{TableBench},
the dark photon $\gamma^\prime_1$ will decay to $e^+e^-$,
thus suffered from the constraints.
We choose the parameter values outside the red region in Fig.~\ref{Fig:Bound} for models $a$ and $b$.
For models $c$-$g$, marked by the shaded light blue in Table~\ref{TableBench},
the dark photon $\gamma^\prime_1$ will mostly decay into dark fermion pairs
and the 511~keV signal constraint does not impose constraints on these models.
We can then choose the parameters within the red region in Fig.~\ref{Fig:Bound} safely for models $c$-$f$.
In model $g$,
the dark photon $\gamma_1^\prime$ possessing a large mass of 170~GeV is also verified to be consistent with experimental data,
though its mass exceeds the range displayed in Fig.~\ref{Fig:Bound} and thus it is not included in the figure.

In the computation of solving the coupled Boltzmann equations Eqs.(\ref{eq:YchiBol})-(\ref{eq:BEjh}),
we take the initial values of $\eta=10,\,\zeta=0.1$
at the temperature $T_{h1}=10^{5}$ GeV.
We take the benchmark models $a$ and $d$ as examples to illustrate the details of the hidden sector evolution.
The full evolution of hidden sector particles as well as the interaction
rates of several important processes among $U(1)$ hidden sectors compared with
the Hubble parameter versus temperature are plotted in Fig.~\ref{Fig:Evo} Panels (c) and (d).

As we can see from Fig.~\ref{Fig:Evo} Panel (c),
for Case 1 model $a$,
the comoving number densities
of the $U(1)^\prime_{1}$ sector particles $\chi_{1},\gamma'_{1}$
accumulate via the freeze-in processes $i\bar{i}\to\chi_{1}\bar{\chi}_{1}$, $i\bar{i}\to\gamma'_{1}$,
and later reach the thermal equilibrium within the $U(1)^\prime_{1}$ sector.
The black vertical dash line marks the temperature at which the
interaction rate of the process $\gamma_{1}^{\prime}\gamma_{1}^{\prime}\to\chi_{1}\bar{\chi}_{1}$
drops below the Hubble expansion rate, which indicates the process $\gamma_{1}^{\prime}\gamma_{1}^{\prime}\to\chi_{1}\bar{\chi}_{1}$
becoming inactive and thus the dark freeze-out occurs in the $U(1)_{1}^{\prime}$ sector,
corresponding to the number density of $\chi_{1}$ going down.
The dark freeze-out ended as the process $\chi_{1}\bar{\chi}_{1}\to\gamma_{1}^{\prime}\gamma_{1}^{\prime}$ becoming inactive,
denoted by the black vertical dash-dotted line,
and the number density of $\chi_{1}$ tends to level off at the observed value of dark matter relic density.
The light blue line in the lower figure shows the thermally averaged decay width of $\gamma_{1}^{\prime}$ to SM fermion pairs
starts to exceed the Hubble expansion rate and since then $\gamma_{1}^{\prime}$ will decay out of equilibrium,
and the number density of $\gamma_{1}^{\prime}$ eventually decline to zero before BBN.
The dark fermion $\chi_{2}$ and the dark photon $\gamma_{2}^{\prime}$ are primarily produced
by the $U(1)_{1}^{\prime}$ sector particles through the interactions
$\chi_{1}\bar{\chi}_{1}\to\chi_{2}\bar{\chi}_{2}$,
$\chi_{1}\bar{\chi}_{1}\to\gamma_{1}^{\prime}\gamma_{2}^{\prime}$,
$\gamma_{1}^{\prime}\gamma_{1}^{\prime}\to\chi_{2}\bar{\chi}_{2}$,
while the freeze-in from SM particles such as $i\bar{i}\to\chi_{2}\bar{\chi}_{2}$,
$i\bar{i}\to\gamma_{2}^{\prime}$ can be safely ignored.
Within the $U(1)_{2}^{\prime}$ sector,
interactions $\chi_{2}\bar{\chi}_{2}\leftrightarrow\gamma_{2}^{\prime}\gamma_{2}^{\prime}$
have important impacts on the final relic abundance of $\chi_2, \gamma_2^\prime$ related to $g_{X_{2}}$.

For Case 2 model $d$ shown in Fig.~\ref{Fig:Evo} Panel (d),
the couplings are set with the relation $\delta_{1} \ll g_{X_{2}}\delta_{2}$,
resulting in most of $\gamma^\prime_{1}$ decaying into the dark fermion $\chi_{2}$ through the intensive process $\gamma^\prime_{1}\to\chi_{2}\bar{\chi}_{2}$,
and thus circumvent the 511~keV signal constraints shown in Fig.~\ref{Fig:Bound}.
In Fig.~\ref{Fig:Evo} Panel (d), the black vertical dash line shows the temperature
at which the thermally averaged decay width of $\gamma'_{1}\to\chi_{2}\bar{\chi}_{2}$
exceeds the Hubble expansion rate and thus this decay process becomes active,
in accordance with a steep drop of the number density of $\gamma^\prime_{1}$.
The production of dark photon $\gamma^\prime_{2}$ is dominated by the interaction
$\chi_{2}\bar{\chi}_{2}\to\gamma_{2}^{\prime}\gamma_{2}^{\prime}$
within the $U(1)^\prime_{2}$ hidden sector.
The dark photon darker matter $\gamma^\prime_{2}$
ultimately constitutes a small fraction of the total dark matter content,
contributing to the 511 keV flux through decay $\gamma'_{2}\to e^{+}e^{-}$.


For all benchmark models we investigate in this work, as shown in Table~\ref{TableBench},
the dark photon $\gamma_{2}^{\prime}$, which constitutes a minor fraction of
the total observed dark matter relic density,
can decay primarily into $e^{+}e^{-}$ pairs,
offering a potential explanation for the observed galactic 511 keV signal.
Using Eqs.~(\ref{eq:flux})-(\ref{eq:rateDecay}), the 511 keV photon flux generated
by the decay of the dark photon $\gamma^\prime_{2}$ is computed to be
\begin{equation} \Phi_{511}=\frac{f_{\gamma'_{2}}}{8\pi}\int\mathrm{d}\Omega\int_{\mathrm{l.o.s}}\frac{\rho\left(r\right)}{M_{\gamma_{2}^{\prime}}}\Gamma_{\gamma'_{2}\to e^{+}e^{-}}\mathrm{d}s\,,
\end{equation}
where the decay width $\Gamma_{\gamma'_{2}\to e^{+}e^{-}}$ is given
by
\begin{align}
	\Gamma_{\gamma_{2}^{\prime}\to e^{+}e^{-}} & \approx\frac{(\delta_{1}\delta_{2}g_{Y}c_{W}^{2}\beta^{2})^{2}}{12\pi}M_{\gamma_{2}^{\prime}}\sqrt{1-\frac{4m_{e}^{2}}{M_{\gamma_{2}^{\prime}}^{2}}}\biggl(1+\frac{2m_{e}^{2}}{M_{\gamma_{2}^{\prime}}^{2}}\biggr)\:,
\end{align}
with $\beta^{2}\equiv M_{\gamma_{2}^{\prime}}^{2}/M_{\gamma_{1}^{\prime}}^{2}$
defined earlier.

We adopt a NFW profile and an Einasto profile, c.f., Eqs.~(\ref{eq:NFW})-(\ref{eq:Einasto}),
with $\gamma=1$, $\alpha=0.17$, $r_{s}=20~\mathrm{kpc}$, and the scale
density $\rho_{s}$ is normalized by the local density $\rho_{\odot}=0.43~\mathrm{\,GeV/cm^{3}}$~\cite{Salucci:2010qr} at the Sun's location $R_{\odot}=8.5~\mathrm{kpc}$.
The bulge flux can be calculated by integrating over the inner part
of the Galaxy ($|l|\lesssim30^\circ$, $|b|\lesssim15^\circ$), or over the Gaussian
full-width-half-maximum (FWHM) corresponding to half of the total
value. In Table~\ref{TableFlux} we list the results of bulge flux for benchmark
model $a$ including two integration range under the NFW and Einasto profile,
which are consistent with the measured bulge flux~\cite{Siegert:2015knp}.
The bulge flux for other benchmark models can also be obtained in the same way.

\begin{table}
	\begin{center}		
		\begin{tabular}{|c|c|c|}
			\hline
		    The bulge flux & $\Phi_{511}^{\mathrm{NFW}}$ & $\Phi_{511}^{\mathrm{Einasto}}$ \tabularnewline
			\hline
		    $|l|\lesssim30^\circ$, $|b|\lesssim15^\circ$ & 8.3 & 9.7 \tabularnewline
			\hline
			FWHM $\simeq20.55^\circ$ & 4.8 & 6.0 \tabularnewline
			\hline
		\end{tabular}
	\end{center}
	\caption{The results of bulge flux for benchmark model $a$ using different
		integral range for the two dark matter density profiles we consider,
        which are consistent with the measured bulge flux~\cite{Siegert:2015knp}.
        All fluxes are in units of $10^{-4}~\mathrm{\,ph\,cm^{-2}\,s^{-1}}$ and
		the FWHM value used here is for the broad bulge.}
	\label{TableFlux}
\end{table}

Thus we find the dark photon darker matter from the two-$U(1)$ model discussed in this paper remains highly promising
in addressing the 511 keV signal consistent with various dark matter density profiles.
Finally we summarize the following relation
\begin{equation}
	\frac{\tau_{X\to e^{+}e^{-}}(\mathrm{sec})\times M_{X}(\mathrm{MeV})}{f_{X}\times\mathrm{Br}(X\to e^{+}e^{-})}\sim10^{\,27}\,,
\end{equation}
for the parameter values we take in the dark matter profiles,
to explain the 511~keV photon signal by the dark matter $X$ decaying to $e^{+}e^{-}$ pairs.

\section{Conclusion}\label{Sec:Con}

In recent years, various dark matter direct and indirect detection experiments,
from earth-based experiments to astrophysical measurements,
impose unprecedented stringent bounds to dark matter models.
The parameter space for a large number of dark matter models are severely constrained.
The relic density $\Omega h^2 = 0.12$ now becomes a constraint to dark matter models
rather than an ultimate goal to achieve.
WIMP dark matter models, which are the most intensively studied from both theoretical and experimental aspects,
are tightly constrained especially from the past decade.
In addition to the traditional secluded dark matter annihilating channels~\cite{Pospelov:2007mp},
another possibility that the dark matter may primarily annihilating to darker concealed sectors are investigated in~\cite{Feng:2024blk}.
Since dark matter has only been detected through gravitational observations,
it is possible that dark matter participates in rather strong interactions within multiple hidden sectors,
while being only ultraweakly coupled to the SM.
An alternative possibility of the dark matter generation is through freeze-in mechanism~\cite{Hall:2009bx,Chu:2011be,Bernal:2017kxu}.
The freeze-in mechanism assumes a negligible initial abundance of the dark particles,
which gradually accumulate as the Universe cools down, from their feeble couplings with the visible sector.

In this paper, we explore the possibility that dark matter,
which does not directly couple to the SM sector,
can be predominantly generated from a frozen-in hidden sector through a further freeze-in process.
Dark matter candidates from hidden sectors indirectly coupled to the SM,
will exhibit weaker-than-ultraweak interactions with the SM sector,
and are thus referred to as ``darker matter''.
As multiple hidden sector may involve more new interactions and thus lead to more complex evolutions,
the study of dark matter from darker hidden sectors,
which indirectly couple to the SM sector, requires a deeper exploration into the dark side of the Universe.
Such investigations also require the method for analyzing the evolution of hidden sector particles and their temperatures,
as developed in~\cite{Aboubrahim:2020lnr}.

To demonstrate the robustness of darker matter candidates produced through a two-step freeze-in process,
we employ a straightforward two-$U(1)$ model for illustration, as shown in Fig.~\ref{Fig:2U1Model}.
We discuss in general the kinetic mixing between two extra $U(1)$ gauge fields and the hypercharge gauge field.
We derive the diagonalization of the mixing matrix by using the perturbation method
and carry out all relevant couplings between hidden sector particles and SM particles.
We compute the full evolution of the two extra $U(1)$ sectors beyond the SM
and discuss several characteristic benchmark models in several distinct cases,
paying particular attention to the darker matter candidates from the darker hidden sector.

The phenomenology of physics from the darker hidden sector is rich and intriguing.
We note that a confirmation of the
velocity-dependence of the darker matter self-interaction cross-section,
as shown in Fig.~\ref{Fig:selfint},
closely matches the observational data ranging from galaxy scales to galaxy cluster scales.
We extensively investigate the dark photon darker matter interpretation of the 511~keV photon signal
and find the two-$U(1)$ model discussed in this work
remains a strong candidate for explaining the 511 keV signal
in concordance with various dark matter density profiles.
Model $y$ from Table~\ref{TableBench_GeV} is of particular interest,
as it demonstrates that the self-interacting darker matter,
which interacts weaker-than-ultraweakly with the SM particles,
can simultaneously account for the cosmic small-scale structure anomalies and the galactic 511~keV photon line signal.

To sum up,
darker matter candidates freeze-in from a frozen-in hidden sector,
couple weaker-than-ultraweakly to the SM particles,
possess intriguing physical properties and have the potential to explain many unresolved problems in the Universe.
The two-step freeze-in process involves more complex interactions beyond the SM
governing the evolution of all dark particles across multiple hidden sectors.
The exploration into the dark side of the Universe may continue to reveal the deeper mysteries of our cosmos.\\

\noindent \textbf{Acknowledgments: }

WZF has benefited from useful discussions with Xiaoyong Chu.
This work is supported by the National Natural Science Foundation of China under Grant No. 11935009.

\appendix

\section{Multi-temperature universe and evolution of hidden sector particles}\label{App:MultiT}

In this Appendix we review the derivation of the evolution of
hidden sector particles and the hidden sector temperature~\cite{Aboubrahim:2020lnr},
for the one hidden sector extension of the SM.
Assuming the hidden sector interacts with the visible sector feebly and hidden sector possess a separate temperature $T_h$
compared to the visible sector temperature $T$ (the temperature of the observed universe).
From Friedmann equations we have
\begin{equation}
\frac{{\rm d}\rho_{h }}{{\rm d}t}+3H\left(\rho_{h }+p_{h }\right)  =j_{h }\,,\label{EOCh}
\end{equation}
where $\rho_h (p_{h})$ is the hidden sector energy density (pressure)
determined by the particles inside the hidden sector, e.g.,
in the single-$U(1)$ extension of the SM, one has $\rho_{h}=\rho_{\gamma^{\prime}}+\rho_{\chi}$ and $p_{h}=p_{\gamma^{\prime}}+p_{\chi}$.

The total energy density in the universe $\rho$ is a sum of $\rho_h$ and  $\rho_v$ (the energy density in the visible sector),
giving rise to the Hubble parameter depending on both $T$ and $T_{h}$
\begin{equation}
H^{2}=\frac{8\pi G_{N}}{3}\big[\rho_{v}(T)+\rho_{h}(T_{h})\big]
= \frac{8\pi G_{N}}{3} \left(\frac{\pi^{2}}{30}g_{{\rm eff}}^{v}T^{4} + \frac{\pi^{2}}{30}g_{{\rm eff}}^{h}T_{h}^{4} \right)\,,\label{hubble}
\end{equation}
where $g_{{\rm eff}}^{v}\,(g_{{\rm eff}}^{h})$ is the visible (hidden) effective degrees of freedom.
Further one can deduct the derivative of $\rho_v,\rho_h$ w.r.t. $T_h$
\begin{align}
\rho\frac{{\rm d}\rho_{h}}{{\rm d}T_{h}}&=\left(\frac{\zeta_{h}}{\zeta}\rho_{h}-\frac{j_{h}}{4H\zeta}\right)\frac{{\rm d}\rho}{{\rm d}T_{h}}\,,\\
\frac{{\rm d}\rho_{v}}{{\rm d}T_{h}}&=\frac{\zeta\rho_{v}+\rho_{h}(\zeta-\zeta_{h})+j_{h}/(4H)}{\zeta_{h}\rho_{h}-j_{h}/(4H)}\frac{{\rm d}\rho_{h}}{{\rm d}T_{h}}\,,
\end{align}
where $\zeta=\frac{3}{4}(1+p/\rho)$. Here $\zeta=1$ is for the radiation
dominated era and $\zeta=3/4$ for the matter dominated universe.
The total entropy of the universe is conserved and thus
\begin{align}
s & =\frac{2\pi^{2}}{45}\big(h_{{\rm eff}}^{v}T^{3}+h_{{\rm eff}}^{h}T_{h}^{3}\big)\,,\label{entropy}
\end{align}
where $h_{{\rm eff}}^{v}\,(h_{{\rm eff}}^{h})$ is the visible (hidden) effective entropy degrees of freedom.

With manipulations of the above equations,
the evolution of the ratio of temperatures for the two sectors $\eta(T_h)=T/T_{h}$ can be derived as
\begin{equation}
\frac{{\rm d}\eta}{{\rm d} T_{h}}=-\frac{A_{v}}{B_{v}}+
\frac{\zeta\rho_{v}+\rho_{h}(\zeta-\zeta_{h})+j_{h}/(4H)}{B_{v} [ \zeta_{h}\rho_{h}-j_{h}/(4H) ]}
\frac{{\rm d}\rho_{h}}{{\rm d}T_{h}}\,,\label{y30}
\end{equation}
with
\begin{equation}
A_{v} =\frac{\pi^{2}}{30}\Big(\frac{{\rm d}g_{{\rm eff}}^{v}}{{\rm d}T}\eta^{5}T_{h}^{4}+4g_{{\rm eff}}^{v}\eta^{4}T_{h}^{3}\Big)\,,\qquad
B_{v} =\frac{\pi^{2}}{30}\Big(\frac{{\rm d}g_{{\rm eff}}^{v}}{{\rm d}T}\eta^{4}T_{h}^{5}+4g_{{\rm eff}}^{v}\eta^{3}T_{h}^{4}\Big)\,.
\end{equation}

The complete evolution of hidden sector particles can be obtained
by solving the combined differential equations of  the evolution equation for $\eta(T_h)$, together with
the modified (temperature dependent) Boltzmann equations for the hidden sector particles, given by
\begin{equation}
\frac{\mathrm{d}Y_{i}}{\mathrm{d}T_{h}}  =-\frac{s}{H}\frac{{\rm d}\rho_{h}/{\rm d}T_{h}}{4\rho_{h}-j_{h}/H}
\sum  \big( - Y_i^2 \langle \sigma v \rangle^{T_h}_{i \bar{i} \to \times \times} + \cdots \big)\,,
\end{equation}
where $Y_i$ is the comoving number density of the $i$th hidden sector particle
and the sum is over various scatter or decay processes involving this particle,
which may depend on $T_h$ (as shown by the example ${i \bar{i} \to \times \times}$) or depend on the visible sector $T$.
The plus (minus) sign corresponds to increasing (decreasing) one number of the $i$th particle.

The derivation of the evolution for
multi-hidden sector extension of the SM was generalized in~\cite{Aboubrahim:2021ycj,Aboubrahim:2022bzk}.
The complete Boltzmann equations for hidden sector particles as well as hidden sector temperatures
are given in Section~\ref{Sec:Evo2U1} for the two-$U(1)$ model.

\section{Derivation of the $4\times4$ rotation matrix in the two-$U(1)$ model}\label{App:2U1dia}

To obtain the couplings of the physical gauge bosons with fermions,
a simultaneous diagonalization of both the kinetic mixing matrix $\mathcal{K}$
and the mass mixing matrix $M_{{\rm St}}^{2}$ can be done
in two steps: (1) Using a non-unitary transformation $K$
to convert the gauge kinetic terms into the canonical form, and $K$
is given by
\begin{equation}
K=\left(\begin{array}{cccc}
c_{2} & 0 & 0 & 0\\
-c_{1}s_{2} & c_{1} & 0 & 0\\
s_{1}s_{2} & -s_{1} & 1 & 0\\
0 & 0 & 0 & 1
\end{array}\right)\,,
\end{equation}
with $c_{1}=1/\sqrt{1-\delta_{1}^{2}}$, $s_{1}=\delta_{1}/\sqrt{1-\delta_{1}^{2}}$,
$c_{2}=\sqrt{1-\delta_{1}^{2}}/\sqrt{1-\delta_{1}^{2}-\delta_{2}^{2}}$,
$s_{2}=\delta_{2}/\sqrt{1-\delta_{1}^{2}-\delta_{2}^{2}}$. This sets
a new gauge eigenbasis
\begin{equation}
KV^{\prime}\equiv V=(D,C,B,A^{3})^{T}\,,\qquad{\rm or}\qquad V^{\prime}=K^{-1}V\,.
\end{equation}
The gauge kinetic terms then become canonical under the new basis
$V^{\prime}$ and the transformation matrix satisfies $K^{T}\mathcal{K}K=\mathbf{1}$.
(2) At the same time, this new basis also changes the gauge boson
mass terms as
\begin{equation}
V^{T}M_{{\rm St}}^{2}V\to(KV^{\prime})^{T}M_{{\rm St}}^{2}(KV^{\prime})=V^{\prime T}K^{T}M_{{\rm St}}^{2}KV^{\prime}\,.
\end{equation}
Now apply an orthogonal transformation to diagonalize the matrix $K^{T}M_{{\rm St}}^{2}K$
such that
\begin{equation}
O^{T}K^{T}M_{{\rm St}}^{2}KO=D^{2}\,.
\end{equation}
The mass terms reduce to
\begin{equation}
V^{\prime T}K^{T}M_{{\rm St}}^{2}KV^{\prime}=V^{\prime T}O(O^{T}K^{T}M_{{\rm St}}^{2}KO)O^{T}V^{\prime}=(V^{\prime T}O)D^{2}(O^{T}V^{\prime})=E^{T}D^{2}E\,,
\end{equation}
where we have defined the physical mass eigenbasis as
\begin{equation}
E\equiv O^{T}V^{\prime}=O^{T}K^{-1}V\,,\qquad{\rm or}\qquad V=KOE\equiv RE\,.
\end{equation}
The kinetic terms will still stay in the canonical form since $V^{\prime T}V^{\prime}=V^{\prime T}OO^{T}V^{\prime}=E^{T}E$.
Thus the transformation $R\equiv KO$ diagonalizes the kinetic and
mass mixing matrices simultaneously, ant it transforms the original
gauge eigenbasis $V^{T}=(D,C,B,A^{3})$ to the mass eigenbasis $E^{T}=(A^\prime_{2},A^\prime_{1},A^{\gamma},Z)$
following the relation
\begin{equation}
V=RE=(KO)E\,.
\end{equation}
Since the mixing parameters $\delta_{1}$ and $\delta_{2}$ are tiny,
in the mass eigenbasis $A^\prime_{1}$($A^\prime_{2}$) is mostly the $U(1)^\prime_{1}$($U(1)^\prime_{2}$)
gauge boson with a tiny fraction of other gauge bosons in the gauge
eigenbasis.

The interaction of gauge bosons verse fermions can be obtained from
\begin{equation}
\mathcal{L}_{{\rm int}}=\left(g_{x}J_{h},0,g_{Y}J_{Y},g_{2}J_{3}\right)V=\left(g_{x}J_{h},0,g_{Y}J_{Y},g_{2}J_{3}\right)RE\,,
\end{equation}
where $J_{Y}$ and $J_{3}$ are the hypercharge current and the neutral
$SU(2)$ current respectively, and in the hidden sector we only have
the fermion current from $U(1)^\prime_{2}$, i.e., $J_{h}=\bar{\chi}\gamma^{\mu}\chi$.
Now the remaining task is to diagonalize the matrix $K^{T}M_{{\rm St}}^{2}K$.
An orthogonal matrix $R_{1}$
\begin{equation}
R_{1}=\left(\begin{array}{cccc}
1 & 0 & 0 & 0\\
0 & 1 & 0 & 0\\
0 & 0 & c_{W} & -s_{W}\\
0 & 0 & s_{W} & c_{W}
\end{array}\right)\,,
\end{equation}
rotates $B,A_{3}$ into $Z_{{\rm SM}},A^\gamma_{{\rm SM}}$. The reduced
mass matrix now reads
\begin{equation}
R_{1}^{T}K^{T}M_{{\rm St}}^{2}KR_{1}=\left(\begin{array}{cccc}
c_{2}^{2}M_{2}^{2}+s_{2}^{2}(c_{1}^{2}M_{1}^{2}+s_{1}^{2}s_{W}^{2}M_{Z}^{2}) & -s_{2}(c_{1}^{2}M_{1}^{2}+s_{1}^{2}s_{W}^{2}M_{Z}^{2}) & 0 & -s_{1}s_{2}s_{W}M_{Z}^{2}\\
-s_{2}(c_{1}^{2}M_{1}^{2}+s_{1}^{2}s_{W}^{2}M_{Z}^{2}) & c_{1}^{2}M_{1}^{2}+s_{1}^{2}s_{W}^{2}M_{Z}^{2} & 0 & s_{1}s_{W}M_{Z}^{2}\\
0 & 0 & 0 & 0\\
-s_{1}s_{2}s_{W}M_{Z}^{2} & s_{1}s_{W}M_{Z}^{2} & 0 & M_{Z}^{2}
\end{array}\right)\,.
\end{equation}
Then we apply a second rotation $R_{2}$
\begin{equation}
R_{2}=\left(\begin{array}{cccc}
1 & 0 & 0 & 0\\
0 & c_{\psi} & 0 & s_{\psi}\\
0 & 0 & 1 & 0\\
0 & -s_{\psi} & 0 & c_{\psi}
\end{array}\right)\,,
\end{equation}
with the angle $\psi$ given by
\begin{equation}
\tan2\psi=\frac{2\delta\sqrt{1-\delta^{2}}\sin\theta_{W}}{1-\epsilon^{2}-\delta^{2}(1+\sin^{2}\theta_{W})}\,.
\end{equation}
Defining $\mathbf{M}^{2}\equiv c_{1}^{2}M_{1}^{2}+s_{1}^{2}s_{W}^{2}M_{Z}^{2}$,
the mass matrix now becomes
\begin{align}
\mathcal{} & \mathcal{M}\equiv R_{2}^{T}R_{1}^{T}K^{T}M_{{\rm St}}^{2}KR_{1}R_{2}\\
 & =\left(\begin{array}{cccc}
c_{2}^{2}M_{2}^{2}+s_{2}^{2}\mathbf{M}^{2} & s_{1}s_{2}s_{\psi}s_{W}M_{Z}^{2}-s_{2}c_{\psi}\mathbf{M}^{2} & 0 & -s_{2}(s_{\psi}\mathbf{M}^{2}+s_{1}s_{W}c_{\psi}M_{Z}^{2})\\
s_{1}s_{2}s_{\psi}s_{W}M_{Z}^{2}-s_{2}c_{\psi}\mathbf{M}^{2} & c_{\psi}^{2}\mathbf{M}^{2}+s_{\psi}^{2}M_{Z}^{2}-2s_{1}s_{\psi}s_{W}c_{\psi}M_{Z}^{2} & 0 & 0\\
0 & 0 & 0 & 0\\
-s_{2}(s_{\psi}\mathbf{M}^{2}+s_{1}s_{W}c_{\psi}M_{Z}^{2}) & 0 & 0 & M_{Z}^{2}
\end{array}\right)\,,\nonumber
\end{align}
with the basis changed to
\begin{equation}
E_{(0)}\equiv R_{2}^{T}R_{1}^{T}V^{\prime}=R_{2}^{T}R_{1}^{T}K^{-1}V\,,
\end{equation}
i.e.,
\begin{equation}
\left(\begin{array}{c}
A_{2(0)}^{\prime}\\
A_{1(0)}^{\prime}\\
A^\gamma_{(0)}\\
Z_{(0)}
\end{array}\right)=\left(\begin{array}{cccc}
\frac{1}{c_{2}} & 0 & 0 & 0\\
\frac{s_{2}c_{\psi}}{c_{2}} & \frac{c_{\psi}+s_{1}s_{\psi}s_{W}}{c_{1}} & s_{\psi}s_{W} & -c_{W}s_{\psi}\\
0 & \frac{s_{1}c_{W}}{c_{1}} & c_{W} & s_{W}\\
\frac{s_{2}s_{\psi}}{c_{2}} & \frac{s_{\psi}-s_{1}s_{W}c_{\psi}}{c_{1}} & -c_{\psi}s_{W} & c_{\psi}c_{W}
\end{array}\right)\left(\begin{array}{c}
D\\
C\\
B\\
A^{3}
\end{array}\right)\,.
\end{equation}
When the off-diagonal elements of $\mathcal{M}$ are much smaller
than the diagonal elements, one can treat off-diagonal elements of
$\mathcal{M}$ as perturbations.
In our analysis,
for the sets of parameters we choose to explain the 511 keV signal,
the differences between each diagonal elements are much greater than
the off-diagonal elements,
thus applying the perturbation method is valid.

Treating the off-diagonal elements of $\mathcal{M}$ as perturbations,
one finds $E_{(1)}\equiv P^{-1}E_{(0)}$, where $P^{-1}$ is the rotation
matrix, i.e.,
\begin{equation}
\left(\begin{array}{c}
A_{2(1)}^{\prime}\\
A_{1(1)}^{\prime}\\
A^\gamma_{(1)}\\
Z_{(1)}
\end{array}\right)=\left(\begin{array}{cccc}
\frac{1}{N_{1}} & \frac{\varepsilon_{12}}{N_{1}} & 0 & \frac{\varepsilon_{14}}{N_{1}}\\
\frac{-\varepsilon_{12}}{N_{1}} & \frac{1}{N_{2}} & 0 & 0\\
0 & 0 & 1 & 0\\
\frac{-\varepsilon_{14}}{N_{1}} & 0 & 0 & \frac{1}{N_{4}}
\end{array}\right)
\left(\begin{array}{c}
A_{2(0)}^{\prime}\\
A_{1(0)}^{\prime}\\
A^\gamma_{(0)}\\
Z_{(0)}
\end{array}\right)\,,
\end{equation}
where we define
\begin{align}
\varepsilon_{12} & =\frac{\mathcal{M}_{12}}{\mathcal{M}_{11}-\mathcal{M}_{22}}\approx\frac{s_{2}M_{1}^{2}}{M_{\gamma^{\prime}_1}^{2}-M_{\gamma^{\prime}_2}^{2}}\,,\label{eps12}\\
\varepsilon_{14} & =\frac{\mathcal{M}_{14}}{\mathcal{M}_{11}-\mathcal{M}_{44}}\approx\frac{s_{2}s_{\psi}(M_{1}^{2}+M_{Z}^{2})}{M_{Z}^{2}-M_{\gamma^{\prime}_2}^{2}}\,.\label{eps14}
\end{align}
and the normalization factors are thus
\begin{equation}
N_{1}  =\sqrt{1+\varepsilon_{12}^2+\varepsilon_{14}^2}\,,\qquad
N_{2}  =\sqrt{1+\varepsilon_{12}^2}\,,\qquad
N_{4}  =\sqrt{1+\varepsilon_{14}^2}\,.
\end{equation}
The first order wave function can written in terms of the original
gauge eigenbasis as
\begin{equation}
E_{(1)}=P^{-1}E_{(0)}=P^{-1}R_{2}^{T}R_{1}^{T}V^{\prime}=P^{-1}R_{2}^{T}R_{1}^{T}K^{-1}V\,,
\end{equation}
or
\begin{equation}
V=KR_{1}R_{2}PE_{(1)}\equiv RE_{(1)}\,.
\end{equation}
Thus the full rotation matrix is given by
\begin{align}
& R=KR_{1}R_{2}P\\
& =\left(\begin{array}{cccc}
c_{2} & 0 & 0 & 0\\
-c_{1}s_{2} & c_{1} & 0 & 0\\
s_{1}s_{2} & -s_{1} & 1 & 0\\
0 & 0 & 0 & 1
\end{array}\right)\left(\begin{array}{cccc}
1 & 0 & 0 & 0\\
0 & 1 & 0 & 0\\
0 & 0 & c_{W} & -s_{W}\\
0 & 0 & s_{W} & c_{W}
\end{array}\right)\left(\begin{array}{cccc}
1 & 0 & 0 & 0\\
0 & c_{\psi} & 0 & s_{\psi}\\
0 & 0 & 1 & 0\\
0 & -s_{\psi} & 0 & c_{\psi}
\end{array}\right)\left(\begin{array}{cccc}
\frac{1}{N_{1}} & \frac{-\varepsilon_{12}}{N_{1}} & 0 & \frac{-\varepsilon_{14}}{N_{1}}\\
\frac{\varepsilon_{12}}{N_{1}} & \frac{1}{N_{2}} & 0 & 0\\
0 & 0 & 1 & 0\\
\frac{\varepsilon_{14}}{N_{1}} & 0 & 0 & \frac{1}{N_{4}}
\end{array}\right)\nonumber
\end{align}
with its entry elements given by
\begin{gather*}
\mathcal{R}_{11}=\frac{1}{N_{1}}c_{2}\,,\qquad \mathcal{R}_{12}= -\frac{\varepsilon_{12}}{N_{2}}c_{2}\,,\qquad
\mathcal{R}_{13}=0\,,\qquad \mathcal{R}_{14}=-\frac{\varepsilon_{14}}{N_{4}}c_{2}\,,\\
\mathcal{R}_{21}=-\frac{c_{1}s_{2}}{N_{1}}+\frac{\varepsilon_{12}}{N_{1}}c_{1}c_{\psi}+\frac{\varepsilon_{14}}{N_{1}}c_{1}s_{\psi}\,,\quad
\mathcal{R}_{22}=\frac{1}{N_{2}}c_{1}c_{\psi}+\frac{\varepsilon_{12}}{N_{2}}c_{1}s_{2}\,,\qquad
\mathcal{R}_{23}=0\,,\qquad\\
\mathcal{R}_{24}=\frac{1}{N_{4}}c_{1}s_{\psi}+\frac{\varepsilon_{14}}{N_{4}}c_{1}s_{2}\,,\qquad
\mathcal{R}_{31}=\frac{s_{1}s_{2}}{N_{1}}+\frac{\varepsilon_{12}}{N_{1}}\left(s_{\psi}s_{W}-c_{\psi}s_{1}\right)-\frac{\varepsilon_{14}}{N_{1}}\left(s_{1}s_{\psi}+c_{\psi}s_{W}\right)\,,\\
\mathcal{R}_{32}=-\frac{1}{N_{2}}\left(s_{1}c_{\psi}-s_{\psi}s_{W}\right)-\frac{\varepsilon_{12}}{N_{2}}s_{1}s_{2}\,,\qquad
\mathcal{R}_{33}=c_{W}\,,\qquad\\
\mathcal{R}_{34}= -\frac{1}{N_{4}}\left(s_{W}c_{\psi}+s_{1}s_{\psi}\right)-\frac{\varepsilon_{14}}{N_{4}}s_{1}s_{2}\,,\qquad
\mathcal{R}_{41}=-\frac{\varepsilon_{12}}{N_{1}}c_{W}s_{\psi}+\frac{\varepsilon_{14}}{N_{1}}c_{\psi}c_{W}\,,\\
\mathcal{R}_{42}= -\frac{1}{N_{2}}c_{W}s_{\psi}\,,\qquad
\mathcal{R}_{43}=s_{W}\,,\qquad
\mathcal{R}_{44}=\frac{1}{N_{4}}c_{W}c_{\psi}\,.
\end{gather*}

\section{Dark matter self-interaction cross-sections}\label{App:SICS}

The dark matter self-interaction cross-sections for $\chi\bar{\chi}\to\chi\bar{\chi}$,
$\chi\chi\to\chi\chi$, and $\bar{\chi}\bar{\chi}\to\bar{\chi}\bar{\chi}$
are obtained from
\[
\frac{{\rm d}\sigma}{{\rm d}\Omega}=\sum_{i=1}^{3}\frac{\overline{|\mathcal{M}_{i}|^{2}}}{64\pi^{2}s}\,,
\]
where for $\chi\bar{\chi}\to\chi\bar{\chi}$
\begin{align}
	\overline{|\mathcal{M}_{1}|^{2}} & =2g_{X}^{4}\biggl\{\frac{t^{2}+u^{2}+8m_{\chi}^{2}s-8m_{\chi}^{4}}{(s-M_{\gamma^{\prime}}^{2})^{2}+\Gamma_{\gamma^{\prime}}^{2}M_{\gamma^{\prime}}^{2}}+\frac{u^{2}+s^{2}+8m_{\chi}^{2}t-8m_{\chi}^{4}}{(t-M_{\gamma^{\prime}}^{2})^{2}}\nonumber\\
	& +\frac{2\bigl[M_{\gamma^{\prime}}^{4}-M_{\gamma^{\prime}}^{2}(s+t)+st+\Gamma_{\gamma^{\prime}}^{2}M_{\gamma^{\prime}}^{2}\bigr]}{\bigl[M_{\gamma^{\prime}}^{4}-M_{\gamma^{\prime}}^{2}(s+t)+st\bigr]^{2}}(u^{2}-8m_{\chi}^{2}u+12m_{\chi}^{4})\biggr\}\,,
\end{align}
and for $\chi\chi\to\chi\chi$, $\bar{\chi}\bar{\chi}\to\bar{\chi}\bar{\chi}$
\begin{align}
	\overline{|\mathcal{M}_{2,3}|^{2}} & =2g_{X}^{4}\biggl\{\frac{s^{2}+u^{2}-8m_{\chi}^{2}(s+u)+24m_{\chi}^{4}}{(t-M_{\gamma^{\prime}}^{2})^{2}}+\frac{t^{2}+s^{2}-8m_{\chi}^{2}(s+t)+24m_{\chi}^{4}}{(u-M_{\gamma^{\prime}}^{2})^{2}}\nonumber\\
	& +\frac{2\bigl[M_{\gamma^{\prime}}^{4}-M_{\gamma^{\prime}}^{2}(u+t)+ut+\Gamma_{\gamma^{\prime}}^{2}M_{\gamma^{\prime}}^{2}\bigr]}{\bigl[M_{\gamma^{\prime}}^{4}-M_{\gamma^{\prime}}^{2}(u+t)+ut\bigr]^{2}}(s^{2}-8m_{\chi}^{2}s+12m_{\chi}^{4})\biggr\}\,,
\end{align}
where $s$, $t$, $u$ are the Mandelstam variables, $g_{X}$ is the
coupling within the hidden sector.

\end{document}